\documentclass[amsmath,amssymb,onecolumn]{aastex631}
\usepackage {subfigure}
\usepackage{rotating}
\usepackage{longtable}
\usepackage {multirow}

\shortauthors{Deng et al.}
\graphicspath{{./}{figures/}}

\begin{document}

\title{On the Formation of Double Neutron Stars in the Milky Way: Influence of Key Parameters}

\author[0000-0002-1398-2694]{Zhu-Ling Deng}
\affil{School of Astronomy and Space Science, Nanjing University, Nanjing 210023, China; lixd@nju.edu.cn }
\affil{Key Laboratory of Modern Astronomy and Astrophysics (Nanjing University), Ministry of Education, Nanjing 210023, China}
\author[ 0000-0002-0584-8145 ]{Xiang-Dong Li }
\affil{School of Astronomy and Space Science, Nanjing University, Nanjing 210023, China; lixd@nju.edu.cn }
\affil{Key Laboratory of Modern Astronomy and Astrophysics (Nanjing University), Ministry of Education, Nanjing 210023, China}
\author[0000-0003-2506-6909 ] {Yong Shao}
\affil{School of Astronomy and Space Science, Nanjing University, Nanjing 210023, China; lixd@nju.edu.cn }
\affil{Key Laboratory of Modern Astronomy and Astrophysics (Nanjing University), Ministry of Education, Nanjing 210023, China}
\author{Kun Xu}
\affil {School of Sciences, Qingdao University of Technology, Qingdao 266525, China}

\begin{abstract}

The detection of gravitational wave events has stimulated theoretical modeling of the formation and evolution of double compact objects (DCOs). However, even for the most studied isolated binary evolution channel, there exist large uncertainties in the input parameters and treatments of the binary evolution process. So far, double neutron stars (DNSs) are the only DCOs for which direct observations are available through traditional electromagnetic astronomy. In this work, we adopt a population synthesis method to investigate the formation and evolution of Galactic DNSs. We construct 324 models for the formation of Galactic DNSs, taking into account various possible combinations of critical input parameters and processes such as mass transfer efficiency, supernova type, common envelope  efficiency, neutron star kick velocity, and pulsar selection effect. We employ Bayesian analysis to evaluate the adopted models by comparing with observations. We also compare the expected DNS merger
rate in the Galaxy with that inferred from the known Galactic population of Pulsar-NS systems. Based on these analyses we derive favorable range of the aforementioned key parameters.

\end{abstract}


\section{Introduction} \label {sec:intro}
Since the discovery of the first chirp of gravitational waves (GWs) from a binary black hole (BH) merger GW 150914 by the Laser Interferometer Gravitational-wave Observatory \citep[LIGO;][]{Aasi2015,Abbott2016}, there have been 90 compact binary merger events detected \citep[][]{Abbott2021b}. The first double neutron star (DNS) merger GW event GW170817 \citep[][]{Abbott2017b} was found to be accompanied by a short gamma-ray burst \citep[SGRB,][]{Abbott2017a} and a kilonova \citep[][]{Abbott2017c}, providing direct evidence for the connection between these phenomena.

The detection of GW events has also stimulated theoretical modelling of the formation and evolution of double compact objects (DCOs). The proposed formation channels for merging DCOs include isolated binary evolution, chemically homogeneous evolution, dynamical interactions in dense environments, hierarchical triple evolution, binary evolution in active galaxy nucleus disks, etc  \citep[see][for recent reviews]{Barack2019,Mandel2022,Mapelli2022}. For the first time, it is possible to
quantitatively compare the merging rate predictions against observations.

However, ever for the most studied isolated binary evolution channel, there exist big uncertainties in the input parameters and in the treatments of the binary evolution processes. For example, a large fraction of merging DCOs have experienced at least one common envelope (CE) evolution phase \citep[][for a review]{Ivanova2013}. Depending on the CE efficiency $\alpha$ and the binding energy parameter $\lambda$ of the donor's envelope, this channel can produce a large range of the merging rates.  Since they can be broadly consistent with the merger rate inferred from GW observations, it is difficult to put stringent constraints on the above mentioned parameters. In particular, some of the models still adopt a constant $\lambda$, which seems physically unrealistic.

So far, DNSs are the only DCOs for which direct observations are available through traditional electromagnetic astronomy.
A viable model predicting the properties of merging DNSs
detectable by GW observatories should also be simultaneously able to account for the Galactic DNS population, although the known 21 DNS pulsars from radio observations represent a very small fraction of all DNSs in the Galaxy. \citet[][]{Phinney1991} and \citet[][]{Narayan1991} tried to infer the DNS merger rates from binary pulsar observations. Since then, there has been remarkable progress in both observation and theory including the growing DNS data and increasingly accurate treatments of the selection effects. And it has enabled comparing theoretical predictions with the observed DNS characteristics, to judge the feasibility of the models and place useful constraints on the model parameters  \citep[][]{Tutukov1993,Portegies1998,Belczy1999,Voss2003,Wang2006,Belczynski2008,Wong2010,Oslowski2011,Andrews2015,Beniamini2016,Belczynski2017,Tauris2017,Vigna2018,Chruslinska2018,Andrews2019,Andrews2019zezas,Chattopadhyay2020,Chu2022}.

In this work, we construct a large number of models for the formation of Galactic DNSs, taking into account various possible combinations of the critical input parameters and processes like mass transfer efficiency, supernova (SN) type, CE efficiency, NS kick velocity, and selection effect of pulsars.  Through both qualitative and quantitative comparison with the Galactic DNSs,  we aim to obtain useful constraints on the parameters and understanding of the binary evolution processes. Here we focus on isolated binary evolution, since most DNSs are in the Galactic field (see below) and dynamical formation of DNSs in star cluster environments does not contribute significantly to the overall merger rate \citep[][]{Phinney1991,Belczynski2018,Fong2019,Ye2019}.



Since the first DNS system PSR B1913+16 was discovered by \citet[][]{Hulse1975}, there are now 21 DNSs known in the Galaxy \citep[see Table 1, part of data are from the ATNF pulsar catalog\footnote{https://www.atnf.csiro.au/research/pulsar/psrcat/},][]{Manchester2005}. Among them, 18 DNSs are in the Galactic field and 3 are in  globular clusters (GCs). The Galactic field DNSs are thought to be predominantly formed through isolated binary evolution \citep{Tauris2017}.

Most of the pulsars in DNSs are the first-born NS, and are observed as (mildly) recycled pulsars. They have accreted more or less mass from the progenitor star of the secondly born NS. Because of mass accretion-induced field decay \citep{Bhattacharya2002}, the recycled pulsars have a weaker magnetic field than normal pulsars, and a longer observable duration in radio \citep[see Fig.~2 of][]{Chattopadhyay2020}. Therefore, we limit our study of the radio characteristic of DNSs to the first-born pulsars.

In this paper, we combine binary population synthesis (BSE) with pulsar evolution to explore the formation of  Galactic DNSs, their radio characteristics, and expected merger rates. The aim of this work is to evaluate the influence of the key parameters by comparing theory with observation.
The remainder of this paper is organized as follows. We introduce isolated binary evolution, the BSE code and the pulsar evolution model in Section 2. We demonstrate the calculated results in Section 3, and explore the influence of the radio selection effect in Section 4.  We present our discussion and conclusions in Section 4.

\begin{deluxetable*}{cccccccccccc}
\tablenum{1}
\tablecaption{Parameters of the detected Galactic DNSs. The spin period $P$, period derivative $\dot{P}$, magnetic field $B$, orbital period $P_{\rm orb}$, eccentricity $e$, mass $M_{\rm psr}$ of the pulsar, mass $M_{\rm com}$ of the companion NS, the total mass $M_{\rm total}$, the characteristic age $\tau_{\rm c}$, and the merging time $\tau_{\rm gwr}$ due to GW radiation. \label{tab:messier}}
\tablehead{
 & & $P$ & $\dot{P}$ & $B$ & $P_{\rm orb}$ & $e$ & $M_{\rm psr}$ & $M_{\rm com}$ & $M_{\rm total}$ & $\tau_{\rm c}$ & $\tau_{\rm gwr}$ \\
 Name & Type & (ms) & ($10^{-18}$) & ($10^{9}$G) & (days) & & ($M_{\odot}$) & ($M_{\odot}$) & ($M_{\odot}$) & (Myr) & (Gyr)}
\startdata
    \multicolumn{12}{c}{Subpopulation (i)}\\
    \hline
    J0737-3039$^{\rm a}$ & recycled & 22.7 & 1.76 & 2.0 & 0.102 & 0.088 & 1.338 & 1.249 & 2.587 & 204 & 0.086  \\
    B1534+12$^{\rm a}$ & recycled & 37.9 & 2.42 & 3.0 & 0.421 & 0.274 & 1.333 & 1.346 & 2.678 & 248 & 2.7 \\
    J1756-2251$^{\rm a}$ & recycled & 28.5 & 1.02 & 1.7 & 0.320 & 0.181 & 1.341 & 1.230 & 2.570 & 443 & 1.7 \\
    J1906+0746$^{\rm a,b}$ & young & 144.1 & 20300 & 530 & 0.166 & 0.085 & 1.291 & 1.322 & 2.613 & 0.1 & 0.31\\
    J1913+1102$^{\rm a}$ & recycled & 27.3 & 0.161 & 0.63 & 0.206 & 0.090 & $<$1.84 & $>$1.04 & 2.88 & 2687 & 0.47 \\
    J1946+2052$^{\rm c}$ & recycled & 17.0 & 0.9 & 1.2 & 0.078 & 0.06 & $<$1.31 & $>$1.18 & 2.50 & 299 & 0.047-0.049 \\
    \hline
    \multicolumn{12}{c}{Subpopulation (ii)}\\
    \hline
    J0453+1559$^{\rm a}$ & recycled & 45.8 & 0.186 & 0.92 & 4.072 & 0.113 & 1.559 & 1.174 & 2.734 & 3901 & 1500 \\
    J1411+2551$^{\rm d}$ & recycled & 62.4 & 0.0956 & 0.77 & 2.616 & 0.1699 & $<1.62$ & $>0.92$ & 2.538 & 10342 & 490-510 \\
    J1518+4904$^{\rm a}$ & recycled & 40.9 & 0.0272 & 0.29 & 8.634 & 0.249 & 1.41 & 1.31 & 2.718 & 23834 & 9200 \\
    J1753-2240$^{\rm a}$ & recycled & 95.1 & 0.970 & 2.7 & 13.638 & 0.304 & - & - & - &  1553 & $>42940$ \\
    J1755-2550$^{\rm a,b}$ & young & 315.2 & 2430 & 270 & 9.696 & 0.089 & - & $>0.4$ & - & 2 & $>25086$ \\
    J1811-1736$^{\rm a}$ & recycled & 104.2 & 0.901 & 3.0 & 18.779 & 0.828 & $<1.64$ & $>0.93$ & 2.57 & 1832 & 1810-1870 \\
    J1829+2456$^{\rm a}$ & recycled & 41.0 & 0.0525 & 0.46 & 1.176 & 0.139 & $<1.38$ & $>1.22$ & 2.59 & 12373 & 59-61 \\
    J1930-1852$^{\rm a}$ & recycled & 185.5 & 18.0 & 18 & 45.060 & 0.399 & $<1.32$ & $>1.30$ & 2.59 & 163 & 554000-572000 \\
    J1325-6253$^{\rm b,e}$ & recycled & 29.0 & 0.048 & 1.18 & 1.816 & 0.064 & $<1.59$ & $>0.98$ & 2.57 & 19180 & 212 \\
    \hline
    \multicolumn{12}{c}{Subpopulation (iii)}\\
    \hline
    J0509+3801$^{\rm f}$ & recycled & 76.5 & 7.93 & 7.8 & 0.380 & 0.586 & 1.34 & 1.46 & 2.805 & 153 & 0.58 \\
    J1757-1854$^{\rm g}$ & recycled & 21.5 & 2.63 & 2.4 & 0.183 & 0.606 & 1.338 & 1.395 & 2.733 & 130 & 0.076 \\
    B1913+16$^{\rm a}$ & recycled & 59.0 & 8.63 & 7 & 0.323 & 0.167 & 1.440 & 1.389 & 2.828 & 108 & 0.3 \\
    \hline
    \multicolumn{12}{c}{Globular Cluster}\\
    \hline
    B2127+11C$^{\rm a}$ & recycled & 30.5 & 4.99 & 3.8 & 0.335 & 0.681 & 1.358 & 1.354 & 2.713 & 97 & 0.22 \\
    J1807-2500$^{\rm a,b}$ & recycled & 4.2 & 0.0823 & 0.18 & 9.957 & 0.747 & 1.366 &1.206 & 2.572 & 805 & 1100 \\
    J0514-4002$^{\rm h}$ & recycled & 4.99 & 0.0007 & 0.019 & 18.8 & 0.89 & 1.25 & 1.22 & 2.473 & 113000 & 470
\enddata
\tablecomments{
$^{\rm a}$ \citet{Tauris2017} and references therein.
$^{\rm b}$ Not a confirmed DNS system--could also be a WD+NS binary.
$^{\rm c}$ \citet{Stovall2018}.
$^{\rm d}$ \citet{Martinez2017}.
$^{\rm e}$ \citet{Sengar2022}.
$^{\rm f}$ \citet{Lynch2018}.
$^{\rm g}$ \citet{Cameron2018}.
$^{\rm h}$ \citet{Ridolfi2019}.
}
\end{deluxetable*}

\section{Methods} \label{sec:style}

\subsection{Binary Population Synthesis}

{We first briefly introduce the two dominant channels of DNS formation by isolated binary evolution \citep{Bhattacharya1991,Tauris2006,Vigna2018}. In the first channel (Channel I),
the evolution begins from a primordial binary with both components more
massive than $8M_{\sun}$, and the mass ratio ($q=$ primary mass/secondary mass) considerably larger than unity.
The primary star evolves more rapidly to overflow its
Roche-lobe (RL) and transfers mass to the secondary. After its envelope is exhausted, the primary evolves to be a helium star, and finally a NS with a SN explosion. The subsequent evolution of the secondary star leads to a CE phase, in which the NS spirals into the envelope of the secondary star. If the NS is able to expel the envelope, the outcome is a NS-helium star binary in a compact orbit. A DNS finally forms if the binary is not disrupted during the second SN. In the second channel (Channel II), the initial mass ratio $q\sim 1$, and the binary experiences a double-core CE phase. This may leave a double helium star binary in a compact orbit. The DNS then evolves from the double helium star binary \citep{Brown1995,Dewi2006,Hwang2015,Andrews2015}.

For the isolated binary evolution, there are some critical physical processes that have not been well understood, such as the stability of mass transfer (MT), the fate of CE evolution, the SN mechanisms and the natal kicks to the NS \citep[e.g.,][]{Andrews2015,Vigna2018,Shao2018b,Chu2022}. They will be discussed in some detail below.
Besides them, before the second SN, there may exist a Case BB MT phase during which the binary comprises a NS and a naked helium star companion \citep{Habets1986}. The Case BB phase also plays an important role in the formation of DNSs \citep{Vigna2018}. During this phase, the first-born NS undergoes mass accretion and is accelerated to short spin period ($\sim 10-100$ ms, depending on the accretion rate and the amount of accreted mass). This may result in an ultra-stripped helium star before the SN, which will finally evolve to be an ONeMg WD or a NS depending on the initial orbital separation and the mass of the helium star \citep{Tauris2013,Tauris2015,Tauris2017}.

After the birth of the DNS, its orbit keeps shrinking by GW radiation \citep{Peters1964,Hulse1975}, finally producing a merger event that may be associated with a SGRB \citep[depending on the assumptions about the DNS masses and other factors,][]{Abbott2017b,Mosta2020,Mandel2022}. The merger time ($\tau_{\rm gwr}$) is determined by the orbital separation, eccentricity and the NS masses. 
\citet{Andrews2019} accordingly split the Galactic field DNSs into three categories: subpopulation (i) have small eccentricities and $\tau_{\rm gwr}$ less than Hubble time; subpopulation (ii)  contain widely separated systems and will not merge within Hubble time; and  subpopulation (iii) have large eccentricities and $\tau_{\rm gwr}$ less than Hubble time (see also Fig.\,1).

We use the BSE code originally developed by \citet{Hurley2002} to calculate the Galactic DNS distribution. We adopt a star formation rate of 3 $M_{\odot}$ yr$^{-1}$, and follow the binary evolution over 12 Gyr.
The primordial binaries consist of a primary star with mass $M_1$ and a secondary star with mass $M_2$ in a circular orbit with separation $a$ (or orbital period $P_{\rm orb}$).
For the distributions of $M_1$, $M_2$ and $a$, we set $n_{\chi}$ grid points of the parameter $\chi$ logarithmically spaced with
\begin{equation}
    \delta {\rm ln}\chi=\frac{1}{{\rm n}_{\chi}-1}({\rm ln}\chi_{\rm max}-{\rm ln}\chi_{\rm min}).
\end{equation}
If a specific binary $i$ evolves to a DNS, then the binary contributes to the DNS population with a rate
\begin{equation}
     R_{\rm i}=(\frac{f_{\rm bin}}{2})(\frac{SFR}{M_{\star}})\Phi({\rm ln} M_1)\phi({\rm ln} M_2)\Psi({\rm ln} a)\delta {\rm ln} M_1\delta {\rm ln} M_2 \delta {\rm ln} a,
\end{equation}
where $f_{\rm bin}$ is the binary fraction (we assume all-stars are initially in binaries, i.e., $f_{\rm bin}=1.0$), $SFR$ is the star formation rate,  and $M_{\star}=0.5M_{\odot}$ is the average mass for all stars; $\Phi$, $\phi$, and $\Psi$ are the distribution functions.

We assume the distribution of the primary masses to follow the initial mass function $\xi(M_1)$ of \citet{Kroupa1993} in the range of $5-40M_{\odot}$.
Thus,
\begin{equation}
    \Phi({\rm ln} M_1)=M_1\xi(M_1).
\end{equation}
For the secondary masses, we assume that they follow a flat distribution between 0 and $M_1$ \citep{Kobulnicky2007}, then,
\begin{equation}
    \phi({\rm ln} M_2)=\frac{M_2}{M_1}.
\end{equation}
The initial orbital separations $a$ are taken to be uniformly distributed in the range of $3- 10^4 R_{\odot}$ in the logarithm \citep{Abt1983}, so
\begin{equation}
    \Psi(\ln a)=k=0.12328.
\end{equation}

Besides the standard treatments of stellar and binary evolution in the BSE code, we modify and update some key processes in the formation of DNSs. They are described as follows.

\subsubsection{Mass loss, mass transfer and CE evolution}
We adopt the fitting formula of \citet{Nieuwenhuijzen1990} to model stellar wind mass loss. For hot OB stars ($T_{\rm eff}>11000$ K) and stripped helium stars, we replace it with the simulated relations by \citet{Vink2001} and \citet{Vink2017}, respectively.

When the primary star evolves to fill its RL, the first MT begins. The accretion efficiency play a vital role in determining the fate of the binary evolution. If the MT is sufficiently rapid, the accreting secondary star will be out of thermal equilibrium, and the responding expansion may lead  the secondary star to fill its own RL, resulting in a contact binary \citep[][]{Nelson2001}. In this situation, if the primary star is a main sequence star, the binary will merge; otherwise, it will enter a CE phase \citep[][]{deMin2013}.

Here we adopt three models for the accretion efficiency in the primordial binaries.

Model I assumes that the mass accretion rate $\dot{M}_2$ of the secondary star is modulated by its rotation rate: $\dot{M}_{2}=\dot{M}_1\times (1-\frac{\Omega_2}{\Omega_{\rm cr}})$, where $\dot{M}_1$, $\Omega_2$, and $\Omega_{\rm cr}$ are the MT rate, the angular velocity of the secondary and its critical value (break-up rate), respectively. The rotation of the star is set to be rigid, and its variation is affected by both accretion and tidal synchronization\footnote{According to \citet{Shao2016}, for narrow binaries, tidal synchronization can prevent the accreting star from reaching critical rotation, and the overall accretion efficiency is $\sim 0.2-0.7$; for wide systems where tidal torques are no longer effective, the accretor quickly spins up to critical rotation and stops accreting, and the overall accretion efficiency is less than 0.2.}.

Model II assumes that half of the transferred mass is accreted by the secondary.

Model III assumes that the transferred mass is always accumulated by the secondary unless its thermal timescale becomes significantly shorter than the mass transfer timescale. So $\dot{M}_2$ is set to be the MT rate multiplied by a factor of min(10$\frac{\tau_{\dot{M}}}{\tau_{\rm KH,2}}$,1) where $\tau_{\dot{M}}$ is the MT timescale and $\tau_{\rm KH,2}$ is the secondary's thermal timescale \citep{Hurley2002}. Rapid mass accretion may push the accretor beyond its thermal equilibrium, causing it to expand and become over-luminous. According to the results of \citet{Shao2014}, $\tau_{\rm KH,2}$ is usually less than $\tau_{\dot{M}}$, which means that the MT is almost conservative.

In all the three models, the excess material is assumed to be ejected out of the binary in the form of isotropic winds, taking away the accretor's specific angular momentum. Thus, the accretion efficiency affects  not only  the orbital evolution of the binary system, but also the stability of MT \citep{Soberman1997}. According to the calculation by \citet{Shao2014}, the maxium $q_{\rm cr}$ for stable MT are $\sim 6$, 2.5, and 2.2 for models I, II and III, respectively. We adopt their results (Fig.~1) of $q_{\rm cr}$ to deal with the MT processes in the primordial binaries.

In the case of NSs accreting from a main-sequence/giant donor star, we use the Eddington limit to constrain the accretion rate, and adopt the maximal mass ratio $q_{\rm cr}=3.5$ for stable MT following previous works \citep{Tauris2000,Podsiadlowski2002,Shao2012,Deng2020}. For NS-Helium star binaries, we assume that the MT becomes dynamically unstable when the orbital period is less than 0.06 day  \citep{Tauris2015}.

To deal with CE evolution, we compare the orbital energy of the binary and the binding energy of the donor's envelope, to evaluate whether the orbital energy is sufficient to expel the envelope \citep{Webbink1984},
\begin{equation}
    \alpha_{\rm CE}\left(\frac{GM_{\rm 1c}M_2}{2a_{\rm f}}-\frac{GM_1M_2}{2a_{\rm i}}\right)=\frac{G(M_1-M_{\rm 1c})M_1}{\lambda r_{\rm L}},
\end{equation}
where $\alpha_{\rm CE}$ is the CE efficiency, $M_{\rm 1c}$ is the mass of the primary's core, $a_{\rm i}$ and $a_{\rm f}$ are the orbital separations of the binary before and after the CE stage respectively, $G$ is the gravitational constant,  $r_{\rm L}$ is the RL radius of the core, and $\lambda$ is the binding energy parameter of the primary's envelope. We let $\alpha_{\rm CE}=0.1$, 0.5, 1, 2, 5 and 10 to evaluate its possible influence. In the case with $\alpha_{\rm CE}>1$ it is assumed that there are extra energy sources (such as recombination energy) used to expel the stellar envelope (values significantly larger than 1 might not reflect real physical process but for illustrative purposes). The binding energy parameter $\lambda$ depends on the stellar evolutionary stage, and we adopt the calculated results of \citet[][]{Xu2010,Xu2010a} and \citet{Wang2016}.

\subsubsection{SN models and natal kicks}
We use the rapid and delayed explosion models proposed by \citet{Fryer2012} and the stochastic explosion model proposed by \citet{Mandel2020} to process the products of core-collapse supernovae (CCSN). In the rapid and delayed SN models, the CO core masses at the SN determines the remnant masses, which are further increased by accreting from the subsequent fallback material. Stars with inital mass $\sim 9\,M_{\odot}$ are thought to produce electron-capture supernovae (ECSN), and the helium core mass $M_{\rm core}$ is traditionally used to discriminate CCSN from ECSN. \citet{Fryer2012} suggested the following criteria: if $M_{\rm core}<1.83\,M_{\odot}$, there will be a degenerate CO core, which ends up as a CO WD; if $M_{\rm core}>2.25\,M_{\odot}$, a non-degenerate CO core is formed, and subsequent thermonuclear burning leads to the formation of an FeNi core, which ultimately collapses to be a NS or BH; if $1.83\,M_{\odot}<M_{\rm core}<2.25\,M_{\odot}$, the star will form a partially degenerate CO core. If the CO core reaches a critical mass $M_{\rm CO,crit}$($=1.08\,M_{\odot}$), it will non-explosively burn into a degenerate ONe core. If the ONe core mass could increase to $M_{\rm  ecs}=1.38\,M_{\odot}$, the star eventually collapses into a NS through an ECSN, or otherwise forms an ONe WD.

However, there are considerable uncertainties in the triggering condition of ECSN, and MT in binaries makes it more complicated \citep[][]{Nomoto1984,Podsiadlowski2004,Woosley2015,Poelarends2017,Doherty2017}. Bearing this in mind,
\citet[][]{Shao2018a} considered three possible helium core mass ranges $M_{\rm ecs}=(1.83-2.25)\,M_{\odot}$, $(1.83-2.5)\,M_{\odot}$, and $(1.83-2.75)\,M_{\odot}$, within which the star may eventually explode in an ECSN. Their results seem to prefer the helium core mass range  $(1.83-2.75)\,M_{\odot}$ for ECSN for the formation of Galactic DNSs.
In this work, we use two possible helium core mass ranges $M_{\rm ecs}=(1.83-2.25)\,M_{\odot}$, and $(1.83-2.75)\,M_{\odot}$ for ECSN.

In addition, Case BB MT may strip most of the envelope of stars before SN explosions \citep[][]{Tauris2013,Tauris2015,Tauris2017}. For such  ultra-stripped supernovae (USSN), we use the recipe in the the rapid model to estimate the compact star mass.

Taking into account all the above factors, we adopt the following five models for SN explosions:

(1) SN A - [rapid CCSN explosion] $+$ [$M_{\rm ecs}=(1.83-2.25)\,M_{\odot}$ for ECSN];

(2) SN B - [rapid CCSN explosion] $+$ [$M_{\rm ecs}=(1.83-2.75)\,M_{\odot}$ for ECSN];

(3) SN C - [delayed CCSN explosion] $+$ [$M_{\rm ecs}=(1.83-2.25)\,M_{\odot}$ for ECSN];

(4) SN D - [stochastic CCSN explosion] $+$ [$M_{\rm ecs}=(1.83-2.25)\,M_{\odot}$ for ECSN];

(5) SN E - [rapid CCSN explosion] $+$ [$M_{\rm ecs}=(1.83-2.25)\,M_{\odot}$ for ECSN] $+$ [USSN].

The natal kick velocities imparted on the NSs are set as follows.

(1) In both rapid and delayed SN  models, the kick velocity is assumed to follow a Maxwellian distribution with the dispersion velocity $\sigma_{\rm k}=150$ or 300 km\,s$^{-1}$ for CCSN, and $\sigma_{\rm k}=20$ or 40 kms$^{-1}$ for ECSN.

(2) The remnant masses and natal kicks in stochastic SN satisfy some specific probability distributions depending on the masses of the CO core \citep{Mandel2020}.

(3) For USSN, we use the formula given by \citet{Andrews2019zezas} to evaluate the kick velocity \citep[the best fit model of the SN kicks in][]{Bray2018}:
\begin{equation}
    V_{\rm k}=|100\frac{M_{\rm He}-M_{\rm NS}}{M_{\rm NS}}-170|\,\rm km s^{-1},
\end{equation}
where $M_{\rm He}$ and $M_{\rm NS}$ are the masses of the pre-SN helium star and the NS, respectively.

\subsubsection{Binary evolution models}
Adopting different MT models, SN models, kick velocity prescriptions, and CE efficiency $\alpha$, we construct 324 binary evolution models, and present the details in Table 2. The name of each model describes the properties of the key parameters. For example,
model MTISNAk20150$\alpha$0.1 means [MT Model I + (SN A) $+$ ($\sigma_{\rm k}=20/150$\,km$s^{-1}$ for EC/CCSN) $+$ ($\alpha_{\rm CE}=0.1$)].
We evolve $10^8$ binary systems and obtain a few to several hundred thousand newborn DNS systems in each run.

\begin{deluxetable*}{ccccccc}
\tablenum{2}
\tablecaption{Parameter settings for binary stellar evolution models}
\tabletypesize{\scriptsize}
\tablehead{
	Model          &   Accretion$^{a}$     & CCSN  & USSN & Kick (Maxwell) & Notes & $\alpha_{\rm CE}^{b}$ \\
  &  efficiency           & Mechanism  &  &  $\sigma_{\rm ECSN},\sigma_{\rm CCSN}$ (km\,s$^{-1}$) &  & }
\startdata
	MTISNAk20150$\alpha$x &   \multirow{18}{*}{$1-\frac{\Omega_2}{\Omega_{\rm cr}}$ }  &  rapid    & no &    20,\,150  &     &   \multirow{18}{*}{0.1-10}    \\
	MTISNAk20300$\alpha$x     &  & rapid   & no &      20,\,300    &   &       \\
	MTISNAk40150$\alpha$x     &   & rapid  & no &      40,\,150   &    &      \\
	MTISNAk40300$\alpha$x     &  & rapid  & no &     40,\,300   &    &       \\
	MTISNBk20150$\alpha$x     &  & rapid   & no &      20,\,150  &  ECSN range$^{c}$   &       \\
	MTISNBk20300$\alpha$x     &  & rapid  & no &     20,\,300  &   ECSN range$^{c}$  &      \\
	MTISNBk40150$\alpha$x     &  & rapid   & no &     40,\,150    & ECSN range$^{c}$  &       \\
	MTISNBk40300$\alpha$x     &  & rapid   & no &     40,\,300    & ECSN range$^{c}$  &       \\
	MTISNCk20150$\alpha$x     &  & delay   & no &     20,\,150    &   &       \\
	MTISNCk20300$\alpha$x     &  & delay   & no &     20,\,300    &   &       \\
	MTISNCk40150$\alpha$x     &  & delay   & no &     40,\,150    &   &       \\
	MTISNCk40300$\alpha$x     &  & delay  & no &     40,\,300    &   &       \\
	MTISNDk20$\alpha$x         &  & stochastic  & no &     $\sigma_{\rm ECSN}=20$ &  kick for CCSN$^d$ &    \\
 	MTISNDk40$\alpha$x         &  & stochastic  & no &     $\sigma_{\rm ECSN}=40$ &  kick for CCSN$^d$  &    \\
        MTISNEk20150$\alpha$x     &  & rapid    & yes &  20,\,150   &   kick for USSN$^e$    &    \\
        MTISNEk20300$\alpha$x     &  & rapid    & yes &  20,\,300   &   kick for USSN$^e$    &    \\
        MTISNEk40150$\alpha$x     &  & rapid    & yes &  40,\,150   &   kick for USSN$^e$    &    \\
        MTISNEk40300$\alpha$x     &  & rapid   & yes &    40,\,300   &  kick for USSN$^e$   &   \\
\hline
        MTIISNAk20150$\alpha$x &   \multirow{18}{*}{0.5} & rapid  & no &      20,\,150   &   &  \multirow{18}{*}{0.1-10} \\
        MTIISNAk20300$\alpha$x     &  & rapid  & no &       20,\,300     &  &       \\
        MTIISNAk40150$\alpha$x     &  & rapid  & no &       40,\,150     &  &       \\
        MTIISNAk40300$\alpha$x     &  & rapid  & no &       40,\,300     &  &       \\
        MTIISNBk20150$\alpha$x     &  & rapid  & no &      20,\,150     & ECSN range$^{c}$ &       \\
        MTIISNBk20300$\alpha$x     &  & rapid  & no &      20,\,300     & ECSN range$^{c}$ &       \\
        MTIISNBk40150$\alpha$x     &  & rapid  & no &      40,\,150     & ECSN range$^{c}$ &       \\
        MTIISNBk40300$\alpha$x     &  & rapid  & no &       40,\,300   &  ECSN range$^{c}$  &      \\
        MTIISNCk20150$\alpha$x     &  & delay  & no &       20,\,150   &    &       \\
        MTIISNCk20300$\alpha$x     &  & delay  & no &       20,\,300   &    &       \\
        MTIISNCk40150$\alpha$x     &  & delay  & no &       40,\,150   &    &       \\
        MTIISNCk40300$\alpha$x     &  & delay  & no &       40,\,300   &    &       \\
        MTIISNDk20$\alpha$x     &   & stochastic  & no &       $\sigma_{\rm ECSN}=20$  &  kick for CCSN$^d$  &      \\
        MTIISNDk40$\alpha$x     &  & stochastic  & no &       $\sigma_{\rm ECSN}=40$   &  kick for CCSN$^d$  &       \\
        MTIISNEk20150$\alpha$x     &  & rapid   & yes &      20,\,150   &  kick for USSN$^e$  &      \\
        MTIISNEk20300$\alpha$x     &  & rapid   & yes &      20,\,300   &  kick for USSN$^e$  &      \\
        MTIISNEk40150$\alpha$x     &  & rapid   & yes &      40,\,150   &  kick for USSN$^e$  &      \\
        MTIISNEk40300$\alpha$x     &  & rapid   & yes &      40,\,300   &  kick for USSN$^e$  &      \\
\hline
        MTIIISNAk20150$\alpha$x & \multirow{18}{*}{min$(10\frac{\tau_{\dot{M}}}{\tau_{\rm KH,2}},1)$}  & rapid  &   no &      20,\,150    &   &  \multirow{18}{*}{0.1-10}    \\
        MTIIISNAk20300$\alpha$x   &  & rapid  & no &       20,\,300   &    &     \\
        MTIIISNAk40150$\alpha$x   &  & rapid  & no &       40,\,150   &    &     \\
        MTIIISNAk40300$\alpha$x   &  & rapid  & no &       40,\,300   &    &     \\
        MTIIISNBk20150$\alpha$x   &  & rapid  & no &       20,\,150   & ECSN range$^{c}$   &     \\
        MTIIISNBk20300$\alpha$x   &  & rapid  & no &       20,\,300   &  ECSN range$^{c}$  &     \\
        MTIIISNBk40150$\alpha$x   &  & rapid  & no &       40,\,150   & ECSN range$^{c}$   &     \\
        MTIIISNBk40300$\alpha$x   &  & rapid  & no &      40,\,300   &  ECSN range$^{c}$  &       \\
        MTIIISNCk20150$\alpha$x   &  & delay  & no &      20,\,150  &    &     \\
        MTIIISNCk20300$\alpha$x   &  & delay  & no &      20,\,300  &    &     \\
        MTIIISNCk40150$\alpha$x   &  & delay  & no &      40,\,150  &    &     \\
        MTIIISNCk40300$\alpha$x   &  & delay & no &       40,\,300 &     &       \\
        MTIIISNDk20$\alpha$x   &  & stochastic  & no &      $\sigma_{\rm ECSN}=20$    & kick for CCSN$^d$  &      \\
        MTIIISNDk40$\alpha$x   &  & stochastic & no &      $\sigma_{\rm ECSN}=40$    & kick for CCSN$^d$  &       \\
        MTIIISNEk20150$\alpha$x   &  & rapid  & yes &       20,\,150  &    kick for USSN$^e$  &     \\
        MTIIISNEk20300$\alpha$x   &  & rapid  & yes &       20,\,300  &    kick for USSN$^e$  &     \\
        MTIIISNEk40150$\alpha$x   &  & rapid  & yes &       40,\,150  &    kick for USSN$^e$  &     \\
        MTIIISNEk40300$\alpha$x   &   & rapid   & yes &     40,\,300  &   kick for USSN$^e$  &     \\
\enddata
\tablecomments{
$^{\rm a}$ The related critical mass ratios ($q_{\rm cr}$) for stable MT comes from \citet{Shao2014} Fig.~1.
$^{\rm b}$ Specifically, all values of $\alpha_{\rm CE}$ are 0.1, 0.5, 1, 2, 5 and 10.
$^{\rm c}$ The ECSN mass range for SN B is 1.83-2.75 $M_{\odot}$.
$^{\rm d}$ Satisfy specific probability distributions described in \citep{Mandel2020}.
$^{\rm e}$ See Eq.~(8).
}
\end{deluxetable*}

Using the formulae of \citet{Peters1964}, we trace the orbital evolution of each DNS with time driven by GW radiation. We limit the evolutionary time to the Galactic age of 12 Gyrs, and calculate the total number of the Galactic DNSs. Each DNS spends specific time in the pulsar phase, and the number and characteristics of observable Pulsar-NS systems are dependent on the pulsar's evolution, the sensitivity of the radio telescopes, and selection effects. They are discussed in the following subsection.

\subsection{Radio characteristics of DNSs}

\citet{Faucher2006} performed Monte Carlo-based population synthesis of the birth properties of Galactic isolated pulsars, their time evolution, and their detection in the Parkes and Swinburne Multibeam surveys. They presented a population model that appears generally consistent with the observations, in which newborn NSs have normally distributed spin periods (with $\mu_{\rm P}=300$ ms and $\sigma_{\rm P}=150$ ms), and log-normally distributed magnetic fields (with $\mu_{\rm log(B/G)}=12.65$ and $\sigma_{\rm log (B/G)}=0.55$). More recently, \citet[][]{Chattopadhyay2020} and \cite{Sgalletta2023} considered several different initial parameter distributions to model the radio characteristics of Galactic DNSs. Their best fit models indicate that the NSs' initial spin periods and magnetic fields are uniformly distributed in the range of $10-100$ ms and $10^{10}-10^{13}$ G, respectively.

The first-born NS in DNSs behaves as radio pulsars mainly after the birth of the second NS. Because it had been (partially) recycled during the CE evolution and the subsequent Case BB MT phase, its spin period and magnetic field at the birth of the second NS (or at the birth of the PSR-NS binary) should deviate from its initial values.
Three-dimensional hydrodynamic simulations by \citet{MacLeod2015} show that NSs undergoing typical CE episodes could accrete a few percent their own mass or less depending on the density gradient inside the envelope and the assumption of NS's hypercritical accretion under neutrino-dominated cooling. The additional phase of MT from the helium star to the NS after the CE stage would also be responsible for the spin and magnetic field of the first-born pulsar \citep[][and references therein]{Jiang2021}. However, the uncertainties in the accreted mass during both processes and the mechanism(s) of accretion-induced magnetic field decay make it difficult to evaluate their influence on the spin and magnetic field evolution in a very reliable way. We instead use the current spin periods and magnetic fields as references for the pulsars. The spin periods of the recycled pulsars in Galactic DNSs are distributed in the range of $17-185.5$ ms (see Table 1). Assuming that the NSs are spinning down due to magnetic dipole radiation, from their spin periods and period derivatives we obtain the magnetic fields  $\sim 10^{8.5}-10^{10.3}$ G. Therefore, we tentatively assume that, at the birth of the second NS, the spin periods of the first-born NSs are uniformly distributed in $10-200$ ms, and the initial magnetic fields (in the logarithm) are uniformly distributed in $8-11$. We note these choices are somewhat arbitrary, and in section 4.3 we vary our assumptions of the spin periods and magnetic fields and discuss their possible influence.

We also assume that, the first NS's surface magnetic field decays spontaneously in an exponential form
\begin{equation}
            B=\left(B_{0}-B_{\min }\right) \times \exp \left(-t / \tau_{\mathrm{d}}\right)+B_{\min },
\end{equation}
where $B_{\rm min}$ is the lowest magnetic field strength taken to be $10^8$ G \citep{Oslowski2011}, and $\tau_{\rm d}$ is the magnetic field decay timescale. We set it to be $10^9$ yr according to the best-fitting models of \citet{Chattopadhyay2020} and \citet{Sgalletta2023}.


We assume that the first-born NS acts as a pulsar and starts spinning down by magnetic dipole radiation after the second NS's birth. We adopt the fitting formula in \citet{Faucher2006} to estimate the pulsar's radio luminosity
\begin{equation}
        \log L=\log \left(L_{0} P^{\epsilon_{\rm P}}{\dot{P}_{15}}^{\epsilon_{\dot{\rm P}}}\right)+L_{\mathrm{corr}}.
\end{equation}
Here, $L_0=0.18\,\rm mJy\,kpc^{-2}$, $\dot{P}=(10^{-15}\,{\rm ss}^{-1})\dot{P}_{15}$ is the period derivative, $\epsilon_{P}=-1.5$, $\epsilon_{\dot{P}}=0.5$, and $L_{\mathrm{corr}}$ is a zero-centered normal distribution with $\sigma_{L_{\rm corr}}=0.8$.
The flux of the pulsar is
\begin{equation}
    F=\frac{L}{4\pi D^{2}},
\end{equation}
where $D$ is the distance between the pulsar and the Solar system. The pulsar could be observed if $F$ is larger than the limiting flux $S_{\rm min}$ of the radio telescope. Approximately half of the Pulsar-NS systems in the Galaxy were detected by the Parkes Multi-beam Pulsar Survey. We set the limiting flux of the Parkes telescope to be 0.2 mJy \citep{Manchester2001}.

The distance between a pulsar and the Solar system depend on the initial location and the subsequent motion of the pulsar under the Galactic gravitational potential. Following \citet{Faucher2006} we set the birth distribution to be
\begin{equation}
    p(r_{\rm birth};{\rm Gauss})\propto \exp[-\frac{(r_{\rm birth}-R_{\rm peak})^2}{2\sigma^2_{r_{\rm birth}}}],
\end{equation}
where $r_{\rm birth}$ is the pulsar's birth distance from the Galactic center, $R_{\rm peak}=7.04$ kpc and $\sigma_{\rm r_{birth}}=1.83$ kpc. Combining the Galactic rotation curve and the kick velocity at the second SN, one can follow the motion of the DNS. We use the {\ttfamily NIGO} code\footnote{Numerical Integrator of Galactic Orbits, which was designed to predict the orbital evolution of test particles moving within a fully-analytical gravitational potential generated by a multi-component galaxy. {\ttfamily NIGO} has been included in the Astrophysics Source Code Library and can be downloaded from https://github.com/LucaJRossi/NIGO \citep{Rossi2015}.} to calculate the three-dimensional (3D) movement of the Pulsar-NS system and the Solar system with time, and to estimate their distance according to their 3D positions. The gravitational potential in the NIGO code includes contributions from the Galactic bulge, the disk, and the dark matter halo \citep{Miyamoto1975,Flynn1996,Navarro1997,Prugniel1997}.

The pulsar's radiation is known to be anisotropic.
The beaming effect significantly influences the observable number of pulsars. We adopt the formula of \citet{Tauris1998} to model the beaming fraction,
\begin{equation}
        f_{\text {beaming }}=0.09\left(\log \frac{P}{10}\right)^{2}+0.03, \quad 0 \leq f_{\text {beaming }} \leq 1
\end{equation}
where $P$ is the spin period of the pulsar in seconds.

Finally, when the pulsars cross a ``death line" in the $P-\dot{P}$ diagram, they stop emitting in the radio band \citep{Rudak1994},
\begin{equation}
        \log _{10} \dot{P}=3.29 \times \log _{10} P-16.55,
\end{equation}
\begin{equation}
          \log _{10} \dot{P}=0.92 \times \log _{10} P-18.65.
\end{equation}
Here, we adopt the second equation (14). Following \citet{Sgalletta2023}, we add a cut-off on the efficiency of radio emission as in \citet{Szary2014} to avoid the accumulation of pulsars at the death lines. The radio efficiency is defined as $\xi=L/\dot{E}$, where  $\dot{E}$ is the pulsar's spin-down power. We set a threshold of $\xi_{\rm max}=0.01$. If $\xi>\xi_{\rm max}$, the pulsar stops emitting radio waves \citep{Szary2014}.


\section{Results}

\subsection{The Orbital period and eccentricity distributions of Galactic DNSs}
\subsubsection{Qualitative analysis}
We plot the distribution of the Galactic DNSs in the $P_{\rm orb}-e$ plane (Figs.~1-4).
Fig.~1 shows the impact of $\alpha_{\rm CE}$ on the number and orbit of DNSs in the SN A model. The number of observable DNSs increases by about 5 times when $\alpha_{\rm CE}$ increases from 0.1 to 10. The modeled DNSs are mostly distributed along the strip extending from lower bottom to upper right (i.e., from near circular short orbits to eccentric wide orbits).  While quite a few observed DNSs are indeed on this strip, most of the model distributions face difficulty in accounting for DNSs with relatively long orbital periods and small eccentricities.

\begin{figure}
    \centering
    \includegraphics[width=17cm]{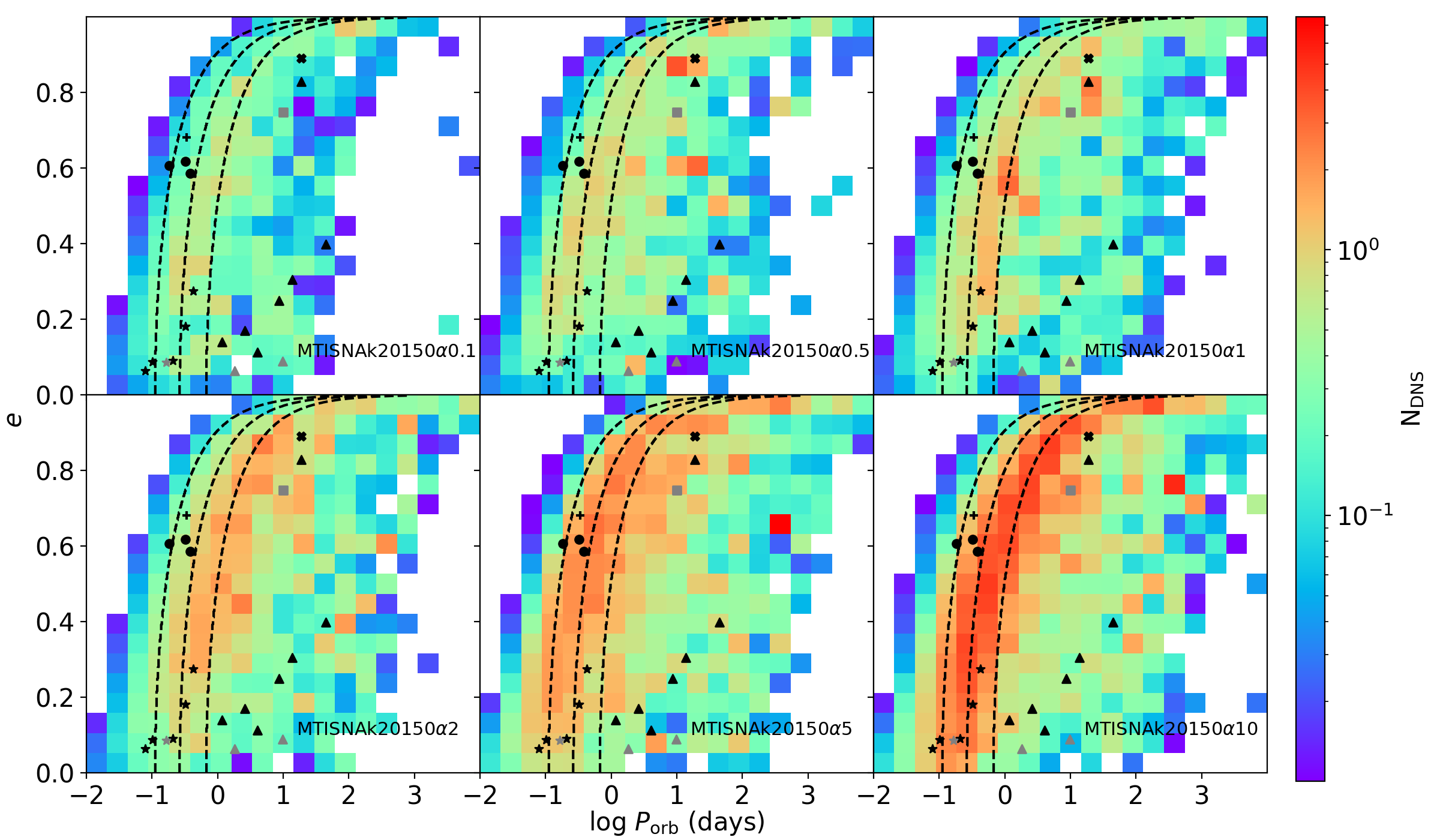}
\caption{Distribution of the Galactic DNSs in the $P_{\rm orb}-e$ plane. Each panel demonstrates a snapshot of the current DNS population in a specific model (MTISNAk40300$\alpha$0.1, MTISNAk40300$\alpha$0.5, MTISNAk40300$\alpha$1, MTISNAk40300$\alpha$2, MTISNAk40300$\alpha$5, MTISNAk40300$\alpha$10). Different colors indicate the predicted DNS numbers, as seen from the color bars on the right. The stars, triangles, and circles represent the Galactic field DNS systems with subpopulation (i), (ii), and (iii), respectively. The square, plus, and cross represent the GC DNSs PSRs J1807-2500, B2127+11C, and J0514-4002, respectively. The gray symbols represent the DNS systems that need further confirmation. The black dash lines represent the merger time of $t_{\rm gwr}=10^8, 10^9,$ and $1.2\times 10^{10}$ yrs, from left to right, respectively. Note that if the observable number is less than 0.01, the color of the pixel grid ($\Delta P_{\rm orb}-\Delta e$) is set to white.}

\end{figure}


Fig.~2 shows the effect of kick velocities in the SN A model with $\alpha_{\rm CE}=1$. The larger kick velocity, the fewer DNSs formed (the observable DNS number in the model with k20150 is approximately four times that in the model with k40300). When the second SN is CCSN, most DNSs are short-period and high-eccentricity systems, and it is difficult (with $\sigma_{\rm k}=150$ km s$^{-1}$) or even impossible (with $\sigma_{\rm k}=300$ km s$^{-1}$) to form long-period systems. When the second SN is ECSN, most DNSs are low-eccentricity and long-period systems, but very few of them can merge within the Galactic age. In summary, both high and low kicks are required for the formation of the Galactic DNSs.
\begin{figure}
    \centering
    \includegraphics[width=17cm]{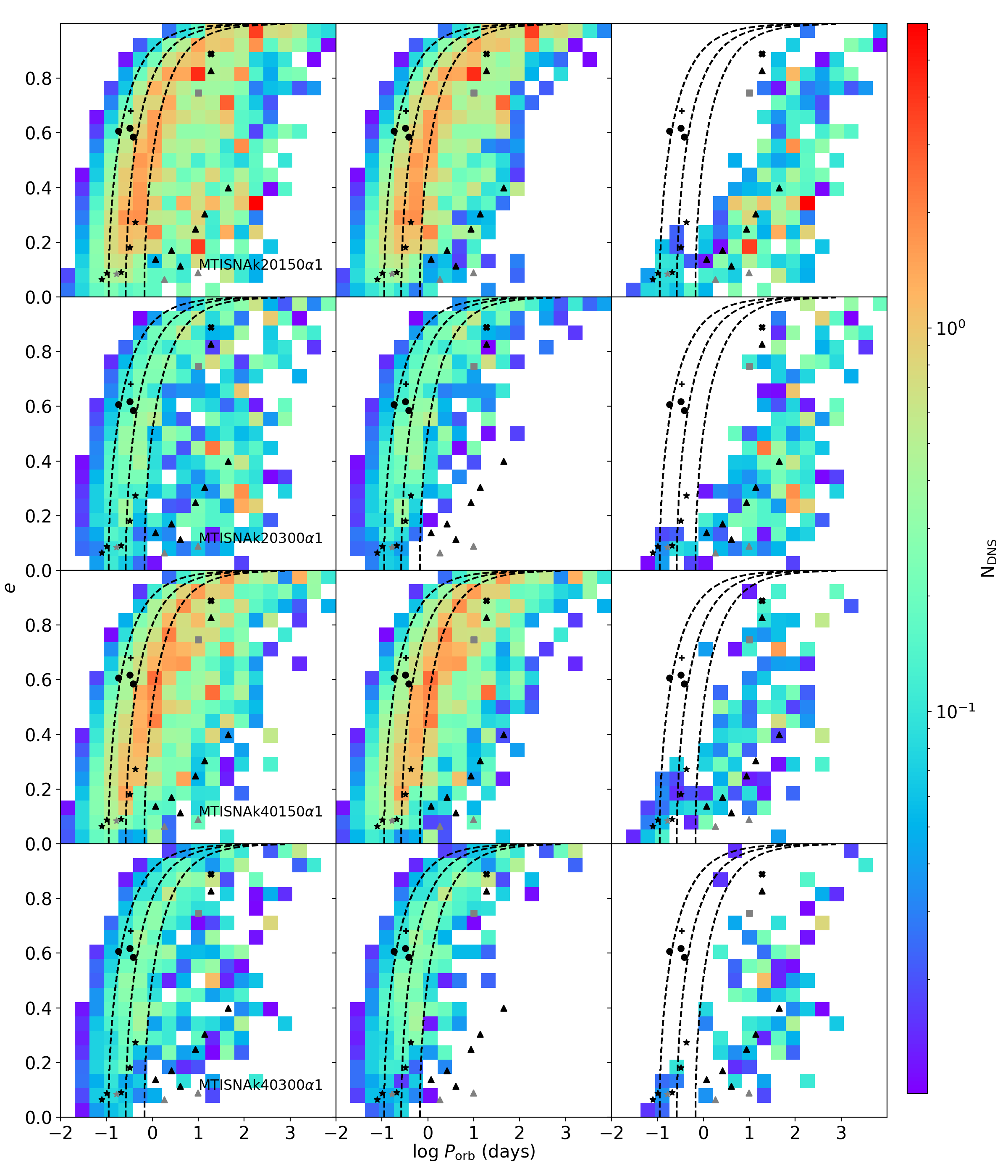}
\caption{Distributions of Galactic DNS binaries in the $P_{\rm orb}-e$ plane under the SN A model with $\alpha_{\rm CE}=1$. The top to bottom panels correspond to models MTISNAk20150$\alpha$1, MTISNAk20300$\alpha$1, MTISNAk40150$\alpha$1, and MTISNAk40300$\alpha$1. The left, middle and right panels represent the distributions of total DNSs, DNSs with the second SN being CCSN and ECSN, respectively.}

\end{figure}

Fig.~3 demonstrates the influence of the initial mass ratio on the formation of DNSs. Similar to \citet{Shao2018a}, both channels I and II contribute to the formation of DNSs. In the SN B model, $\sim 60 \%$ DNSs form through channel I (especially for near circular short orbit DNSs), while in the SN A model, the numbers of DNSs formed through channels I and II are roughly comparable.
\begin{figure}
    \centering
    \includegraphics[width=17cm]{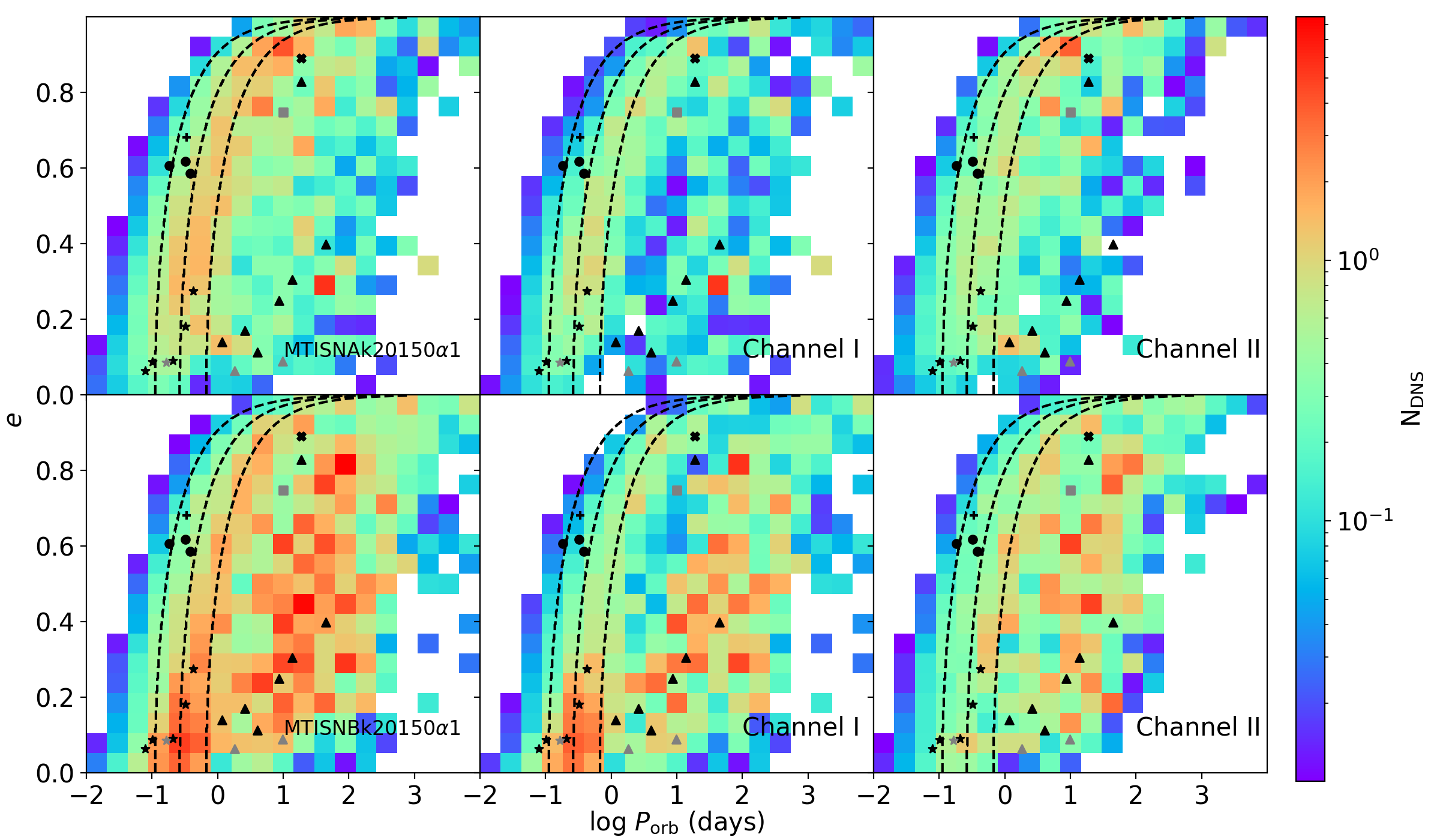}
\caption{Distributions of Galactic DNS systems. The top and bottom panels correspond to models MTISNAk20150$\alpha$1 and MTISNBk20150$\alpha$1, respectively. The left, middle, and right panels represent the distributions of total DNSs, DNSs formed through channel I (high initial mass ratio), and channel II (initial mass ratio close to 1), respectively.}
\end{figure}

Fig.~4 compares the results in different SN and MT models. SN B model produces approximately two times DNSs than other models, and their distribution seems more consistent with observations, especially with MT II. The main reason is that in this model $M_{\rm ecs}=(1.83-2.75)M_{\odot}$, which substantially increases the occurrence rate of ECSN. The relatively low kicks during the SN reduces the disruption probability of  the binaries, and helps form DNSs with low-eccentricity. SN A and C models show similar distributions. The DNSs formed in SN D model are mostly concentrated in the region of $\log P_{\rm orb}(\rm day)\sim -1- 0.5$ and $e\sim 0.2- 0.6$. Compared with SN A model, SN E model includes Case BB MT and USSN, but the results are similar. For given SN prescription,
we find that MT II model produces approximately two times DNSs than MT I model, and approximately four times than MT III model.
%
%
%
\begin{figure}
    \centering
    \includegraphics[width=17cm]{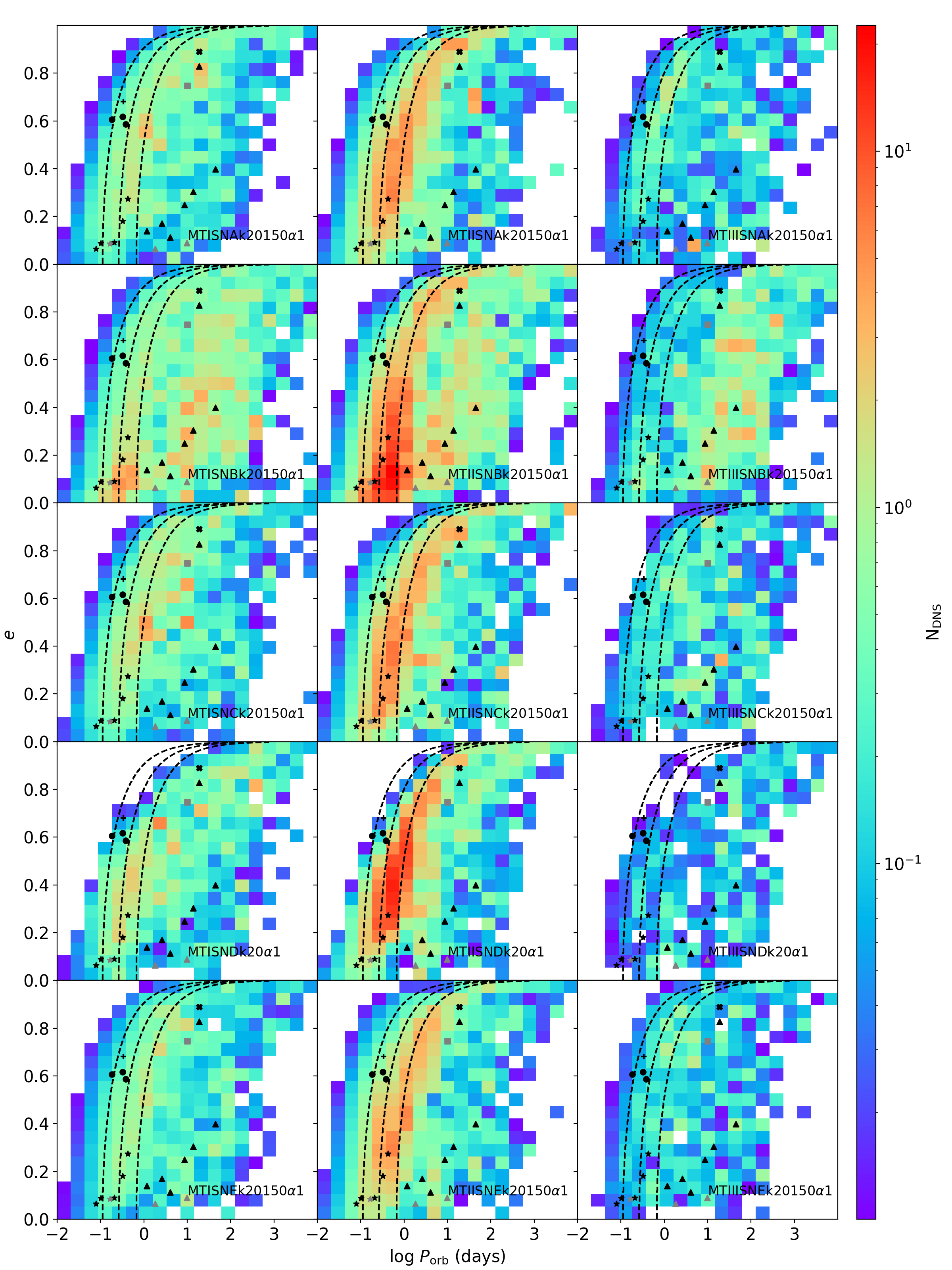}
\caption{Distributions of Galactic DNS systems in different SN and MT models. The top to bottom rows correspond to SN A-E models, and the left, middle and right columns MT I-III models, respectively.}
\end{figure}

%
%

\subsubsection{Quantitative analysis}
We use the Bayesian analysis method employed by \citet{Andrews2015} to quantitatively compare the computed results with observations. We consider two independent orbital parameters $P_{\rm orb}$ and $e$ for each model. According to the Bayesian theory, the posterior probability related to model $M$ given the observed data $D$ is
\begin{equation}
    P(M|D)=\frac{P(D|M)P(M)}{P(D)},
\end{equation}
where $D$ denotes $P_{\rm orb}$ and e, $P(D|M)$ is the likelihood of observed data given that the model is true, and we set it to be $\Lambda(D)$, $P(M)$ is the prior probability of a specific model, and $P(D)$ is a normalizing constant which is independent of the model. Since all models have the same prior probability, $P(M)/P(D)$ can be normalized as a constant $C$. Thus, $P(M|D)$ is  rewritten to be
\begin{equation}
   P(M|D)=C\Lambda (D).
\end{equation}
The relative probability of each model is determined by $\Lambda (D)$ because $C$ is independent of the models. Since the observed DNS systems are independent, the relative probability is equal to the probabilities of each independent system,
\begin{equation}
    \Lambda (D)=\Lambda\left({\log}\,P_{\text {orb }}, e\right)=\prod_{i} P\left({\log}\,P_{\text {orb}, i}, e_{i} \mid M\right).
\end{equation}
Here $P({\rm log}\,P_{\rm orb,i},e_{\rm i}|M)$ is the probability density of a model at a specific point in the ${\log}\,P_{\rm orb}-e$ parameter space. Similar as \citet[][]{Chu2022}, for each pixel ($\log\,P_{\rm orb,i},e_{\rm i}$), we adopt a 2D normal distribution $N({\log}\,P_{\rm orb},e|{\rm log}\,P_{\rm orb,i},e_{\rm i},\sigma^2_{{\log}\,P_{\rm orb,i}},\sigma^2_{e_{\rm i}})|_{M}$ to avoid zero probability at some pixels. The standard deviations of $\log\,P_{\rm orb,i}$ and $e_{\rm i}$ are constant and assigned to be $\sigma_{{\log}\,P_{\rm orb,i}}=0.3$ and $\sigma_{e_{\rm i}}=0.03$.

So far, 18 DNSs have been observed in the Galactic field, of which 3 DNSs have not been fully confirmed (see Table 1). Therefore, we consider two sets of data. The first set of data $D_1$ consists of 15 fully confirmed DNSs, and the second set $D_2$ includes all 18 DNSs. Fig.~5 shows the values of $\Lambda (D_1)$ as a function of the CE parameter $\alpha_{\rm CE}$. Larger value of $\Lambda (D)$ means that the model predictions are more consistent with observations.

\begin{figure}
    \centering
    \includegraphics[width=12cm]{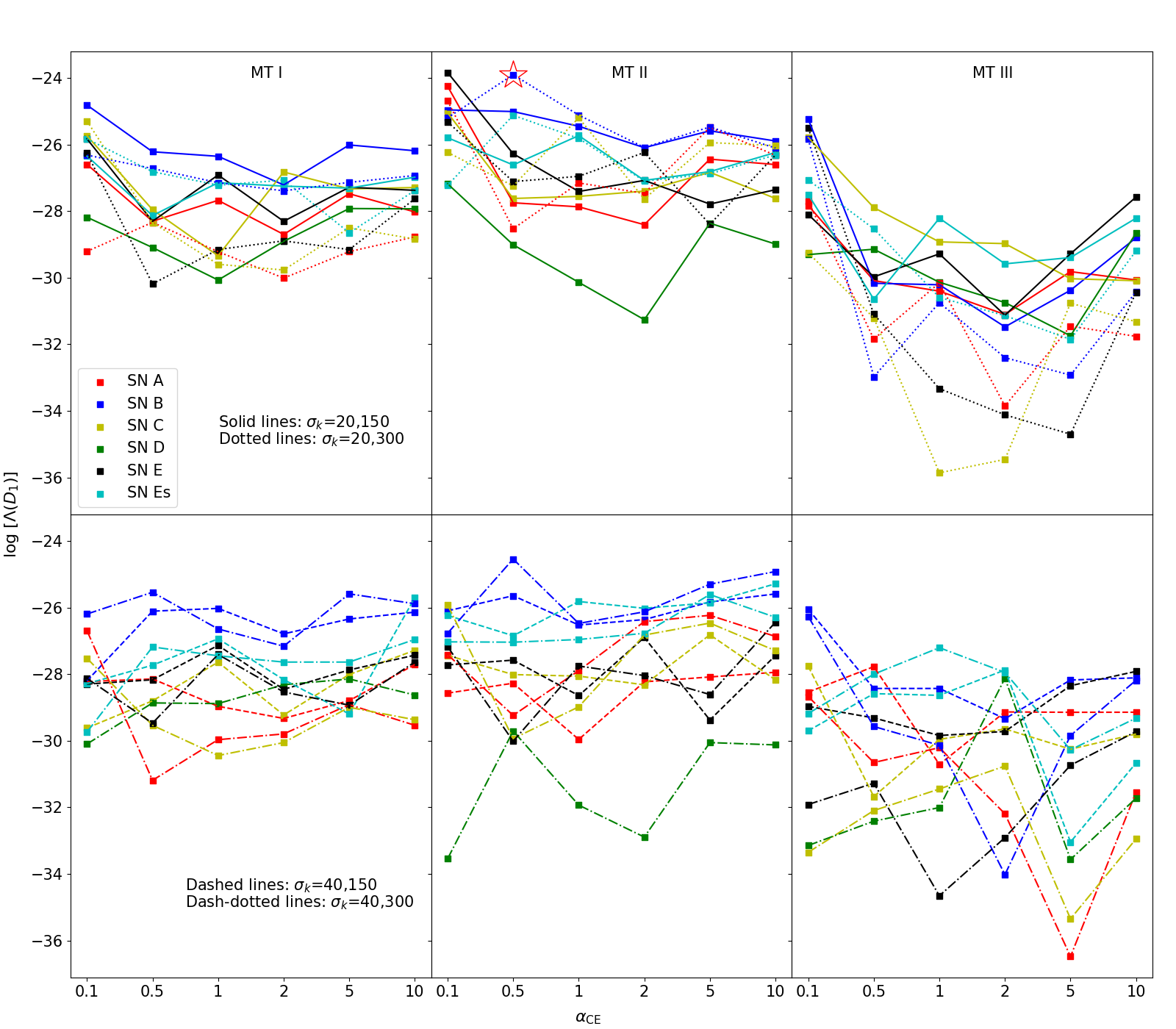}
    \caption{The posterior probability for different models. The red, blue, yellow, green and black colors correspond to SNA-E models, respectively. The solid, dotted, dashed and dash-dotted lines represent the cases with $\sigma_{\rm ECSN, CCSN}$=20/150, 20/300, 40/150 and 40/300 km\,s$^{-1}$. Note that for SN D and E models, the kick velocity distribution is not a single Maxwell distribution (see Table 2). The cyan color corresponds to SNEs. Based on SNE and assuming that the kicks of both USSN and ECSN follow the same distribution (i.e., $\sigma_{\rm USSN}=20/40$ km\,s$^{-1}$). The left, middle, and right panels represent MT I, II, and III models, respectively. The best fit model is marked by a red star. }

\end{figure}

We draw the following conclusions from Fig.~5:

\begin{itemize}
    \item The values of $\Lambda (D)$ vary in a wide range with different input parameters. There is not apparent dependence of $\Lambda (D)$ on the CE efficiency $\alpha_{\rm CE}$ and the kick velocity distribution.
    \item Overall nonconservative MT I and II models are considerably better than nearly conservative MT III models.
    \item SN B model (rapid CCSN explosion and $M_{\rm ecs}=[1.83-2.75]\,M_{\sun}$ for ECSN) is more preferred compared with other SN models.
\end{itemize}

Table A (in Appendix A) shows the results of all the adopted models. We also present the ranking numbers  (level$_1$ and level$_2$) of each model according to their calculated $\log\,\Lambda (D_1)$ and $\log\,\Lambda (D_2)$, as well as the birth rates and merger rates of DNSs.
The top ten models in level$_1$ are MTIISNEk20150$\alpha$0.1, MTIISNBk20300$\alpha$0.5, MTIISNAk20150$\alpha$0.1, MTIISNBk40300$\alpha$0.5, MTIISNAk20300$\alpha$0.1, MTISNBk20150$\alpha$0.1, MTIISNBk40300$\alpha$10, MTIISNBk20150$\alpha$0.1, MTIISNBk20150$\alpha$0.5 and MTIISNCk20150$\alpha$0.1.

\subsection{The merger rate of Galactic DNSs}

Another way to calibrate the models of the formation and evolution of DNSs is comparing the expected DNS merger rate in the Galaxy by first principles with that inferred from the known Galactic population of Pulsar-NS systems.
Figure 6 depicts the DNS merger rates calculated for different models. They are in the range of $\sim 1-70$ Myr$^{-1}$ depending on the model parameters.
In general, the merger rate increases with the CE efficiency $\alpha_{\rm CE}$ and decreases with the kick velocity. The reason is that larger $\alpha_{\rm CE}$ results in more DNSs, and larger kick velocity causes more binary systems to be disrupted. However, Figure 6 also shows that the DNS merger rate does not change monotonically with  $\alpha_{\rm CE}$. In some cases the merger rate decreases with increasing $\alpha_{\rm CE}$. This is because larger $\alpha_{\rm CE}$ results in wider post-CE binaries, which are more easily disrupted after the SN.
Additionally, the merger rate is critically dependent on the MT models. The overall merger rates in MT I and II models are higher than in MT III model.

On the observational side, \citet[][]{Kim2003} developed a Bayesian statistical method to derive the merger rate probability distribution by modelling the Galactic DNS population as well as the selection effects based on the observed properties of known binaries and survey characteristics. Based on this work, \citet[][]{Kalogera2004} estimated the merger rate $\sim 90$ Myr$^{-1}$, considering PSRs B1913+16, B1534+12, and J0737$-$3039A. Considering PSRs B1913+16, B1534+12, J0737$-$3039A, J1756$-$5521, and J1906+0746
with estimated beaming correction factors,
\citet[][]{O'Shaughnessy2010} obtained the merger rate $\sim 60$ Myr$^{-1}$. Taking into account the known DNS binaries with the best observational constraints, including both PSRs J0737$-$3039A and B, \citet[][]{Kim2015} derived the merger rate to be $21^{+28}_{-14}$ Myr$^{-1}$ at 95 per cent confidence. The method was  recently applied by \citet[][]{Pol2019,Pol2020} with updated DNS data, including the highly eccentric DNS system PSR J0509+3801 \citep[][]{Lynch2018}, to estimate the merger rate to be $37^{+24}_{-11}$ Myr$^{-1}$ at 90 per cent confidence. \citet[][]{Grunthal2021} revisited this problem taking account of recent observation of the longitudinal and latitudinal beam shape of PSR J1906+0746 \citep[][]{Desvignes2019}, and derived a Galactic DNS merger rate of $32^{+19}_{-9}$ Myr$^{-1}$ at 90 per cent confidence.

Fig.~6 compares the calculated and inferred DNS merger rates. For MT I and III models, to be compatible with these estimates requires $\alpha_{\rm CE}\sim 5-10$, while for MT II models, the allowed values of $\alpha_{\rm CE}$ reside in a wide range of $0.5-10$. Among the ten best models selected by the Bayesian method in the Section 3.1.2, only models MTIISNBk20300$\alpha$0.5, MTIISNBk40300$\alpha$0.5, and MTIISNBk40300$\alpha$10  match the derived DNS merger rate. So, we conclude that the best fit model is MTIISNBk20300$\alpha$0.5.

\begin{figure}
    \centering
    \includegraphics[width=12cm]{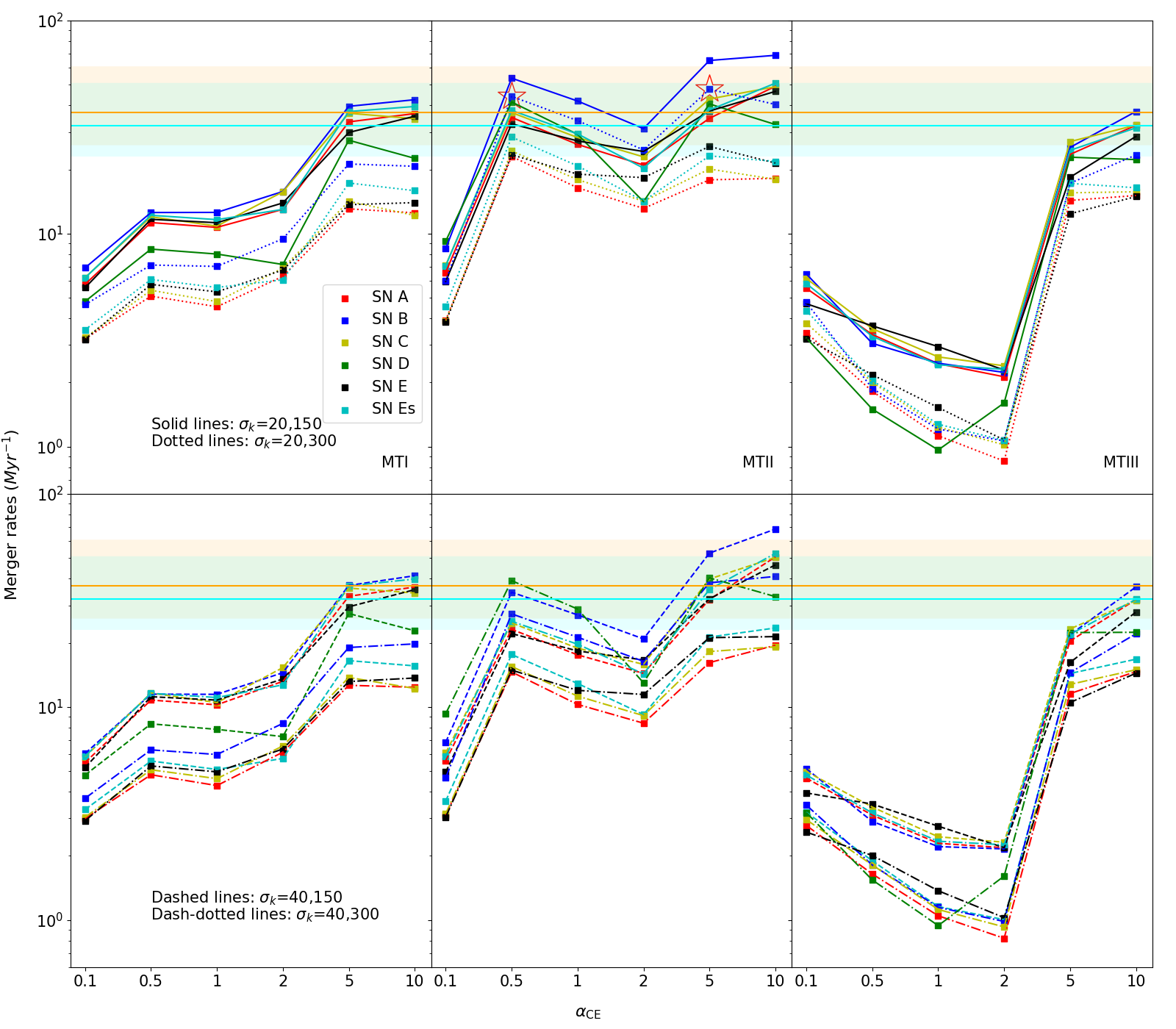}
\caption{Same as Figure 5 but for DNS merger rates. The orange and cyan horizontal lines and the shaded areas represent the merger rates of $37^{+24}_{-11}$\,Myr$^{-1}$ \citep[][]{Pol2019,Pol2020} and $32^{+19}_{-9}$\,Myr$^{-1}$ \citep[][]{Desvignes2019}, respectively. Note that the two red hollow stars  represent the best fit model obtained by adopting and not adopting the Doppler shifting effect.}
\end{figure}
\section{Influence of The Doppler Shifting Effect}

In the above section we have neglected the possible influence of the Doppler shifting effect and assumed that all Pulsar-NS binaries have the same detection probability. Actually it is challenging to detect eccentric binary pulsars with short orbital periods in pulsar searches considering the Doppler modulation of the pulsed signal as the pulsar moves with respect to the center of mass of the binary system during an observation \citep{Johnston1991,Bagchi2013}.
\citet{Johnston1991} developed an analytical framework for calculating the reduction in signal-to-noise ratio caused by binary motion in the case of circular orbits. \citet{Bagchi2013} generalized the work of \citet{Johnston1991} and presented analytical
expressions to calculate the sensitivity reduction for orbits of arbitrary
eccentricity. Based on the results of \citet{Bagchi2013},
\citet{Chattopadhyay2021} derived a fitting formula in the case of a 1.4 $M_{\odot}$ NS and an observation duration of 1000 s at an inclination angle of 60$^{\circ}$, and fit eccentricities via linear regression for $e=0.1$, 0.5, and 0.8. These authors suggested that binary pulsars are detectable if
\begin{equation}
        P_{\rm orb}({\rm day})> m\times P({\rm s}) +c ,
\end{equation}
where
\begin{equation}
        m=m_{\rm m}\times e+c_{\rm m},
\end{equation}
and
\begin{equation}
        c=m_{\rm c}\times e+c_{\rm c},
\end{equation}
where $m_{\rm m}=-8.9$, $c_{\rm m}=-27.68$, $m_{\rm c}=-3.4$, and $c_{\rm c}=-5.72$.

The above criterion depends on both the orbital evolution and the pulsar's spin evolution. The latter is determined by the NS spin period and magnetic field when it behaves as a radio pulsar. In section 2.2 we simply adopt uniform distribution for both $P_{\rm s}$ and $\log B$ inferred from their current observational data. However,
due to previous accretion during the CE stage and the Case BB MT stage, both parameters should be related to the orbital period and eccentricity of the DNS systems\footnote{The wider the initial orbit of the NS-helium star system, the more evolved the helium star by the time it fills its RL and the shorter Case BB MT duration. As a result, the recycling of the first-born NS is less efficient in wider systems, and its rejuvenated spin period may be longer \citep{Tauris2015,Tauris2017}.}. Moreover, the birth\footnote{Here ``birth" means the values at the birth of the DNSs.} $P_{\rm orb}$ and $P_{\rm s}$ of DNS systems are not be exactly reflected by their currently observational data due to the loss of rotational energy and the decay of orbits caused by GW radiation.  Taking these effects into account in a qualitative manner and placing particular weight on the very wide-orbit DNS PSR J1930$-$1852 \citep{Swiggum2015}, \citet{Tauris2017} obtained a simple empirical correlation between their birth $P_{\rm s}$ and $P_{\rm orb}$,

\begin{equation}
P_{\rm s}=(36\pm 14)\,{\rm ms}\,P_{\rm orb}(\rm{day})^{0.4\pm 0.1}.
\end{equation}
We adopt above equation to evaluate the birth spin periods of the first-born pulsars.

As to the pulsar's birth magnetic field, we select ten relatively young DNSs from Table 1 with the characteristic ages less than $10^9$ yr, to ensure that their magnetic fields have not substantially decayed since the birth of the DNSs.
Assuming that the characteristic age reflects the true age, we use the GW-induced orbital evolution functions and a Monte Carlo method to obtain the birth $P_{\rm orb}$ and $e$. Then we employ a nonlinear least squares method to derive the correlation between the magnetic field and the orbital period and eccentricity (see Fig.~7),
\begin{equation}
\log B ({\rm G}) = 1.47 e^{0.051} + 8.34P_{\rm orb}({\rm day})^{0.018}.
\end{equation}
The coefficient of determination of the above fit is 0.764.
\begin{figure}
    \centering
    \includegraphics[width=8cm]{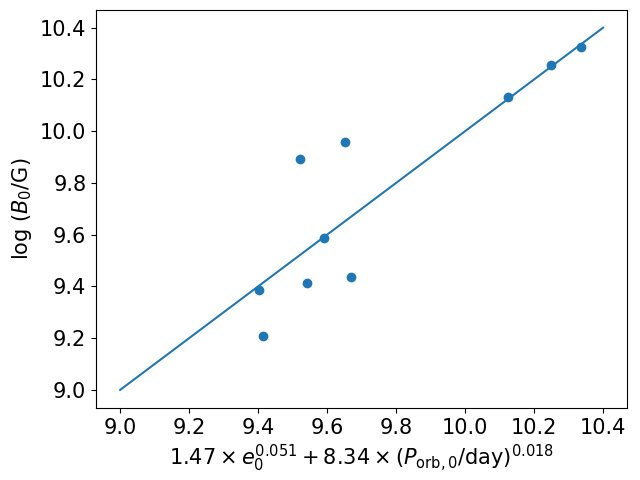}
    \caption{The birth magnetic field ($B_0$) as a function of birth orbital period ($P_{\rm orb,0}$) and birth eccentricity ($e_0$) for selected first-born (recycled) NSs in DNS systems. The derived selected DNS birth data are plotted with blue points. The blue line represents a fit correlation (Eq.~[22]).}
\end{figure}

Note also that the results of \citet{Bagchi2013} suggest a stiff cut-off of radio pulsar detection, which means that no pulsar can be detected if Eq.~(18) is not satisfied. We instead revise it to be a soft cut-off: if a binary crosses the cut-off line, the observable number is reduced by a factor
\begin{equation}
f_{\rm d}=\frac{P_{\rm orb}({\rm day})}{m\times P({\rm s}) +c}.
\end{equation}
According to the results of \citet{Bagchi2013}, we find that $f_{\rm d}$ is limited in the range of $0.3-0.55$.

Fig.~8 compares the cumulative distributions of the orbital period ($P_{\rm orb}$),  eccentricity ($e$), magnetic field strength ($B$), and spin period ($P_{\rm s}$) (from top left to bottom right) of DNSs for both observational and modeled DNSs. We find that considering a stiff cut-off due to the Doppler shifting effect significantly reduces the number of DNS systems with $P_{\rm orb}<1$ day, inconsistent with observations, and the modeled distributions of $e$, $P_{\rm s}$ and $B$ also deviate from observations. On the other hand, the predicted orbital distribution of DNSs without considering the Doppler shifting effect is incompatible with observations. In comparison, when soft cut-off considered, the distributions of all the four parameters seem to better match observations.

\begin{figure}
    \centering
    \includegraphics[width=10cm]{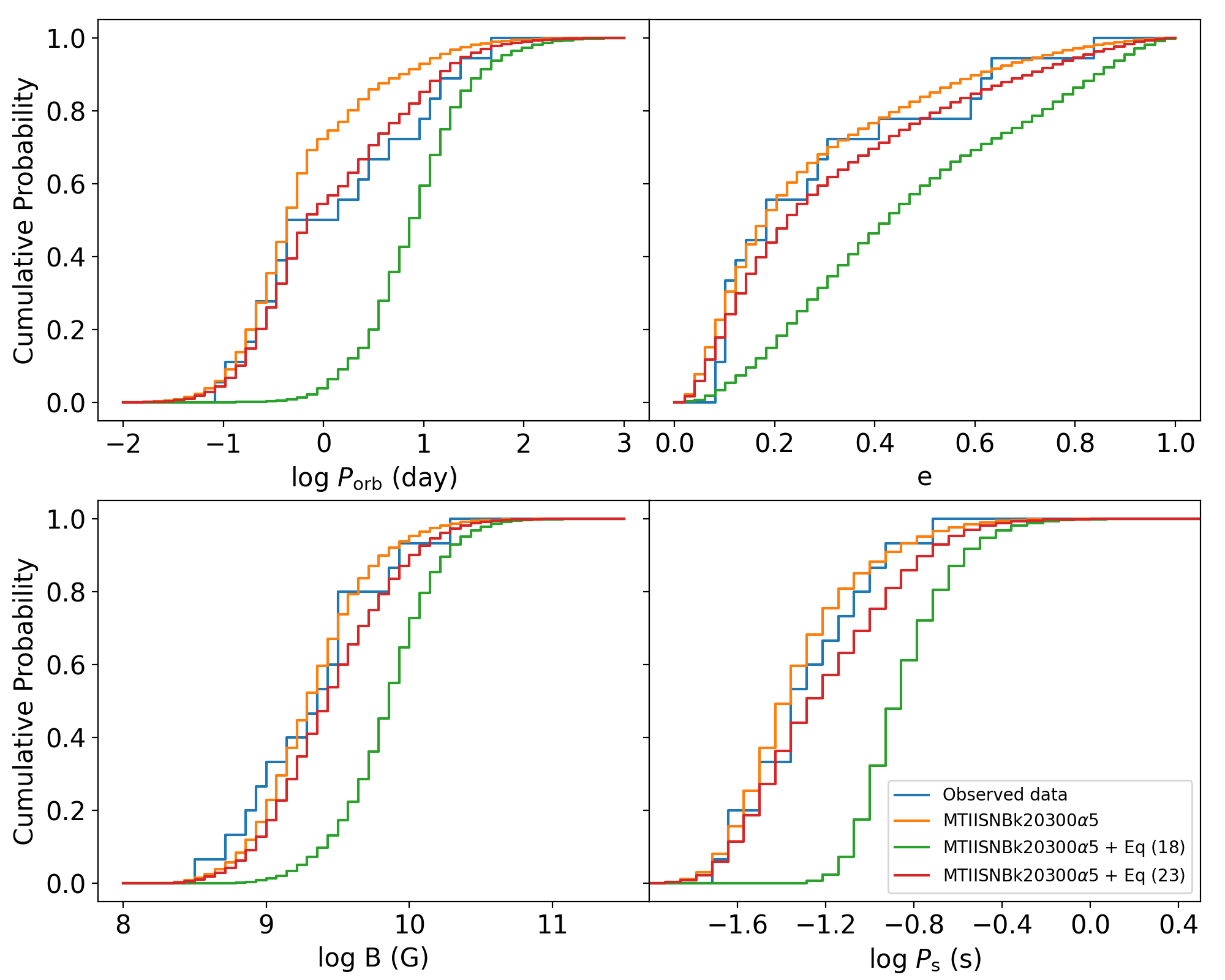}
    \caption{Cumulative distributions of the DNS parameters $P_{\rm orb}$, $e$, $B$ and $P_{\rm s}$ (from top left to bottom right) for the model of MTIISNBk20300$\alpha$5. The blue line represents the observational data, The orange, green, and red lines represents the calculated DNS population that does not take into account the Doppler Shifting effect, adopts hard (Eq.~[18]) and soft (Eq.~[23]) cut-off, respectively.}
\end{figure}

\begin{figure}[htb]
    \centering
    \includegraphics[width=12cm]{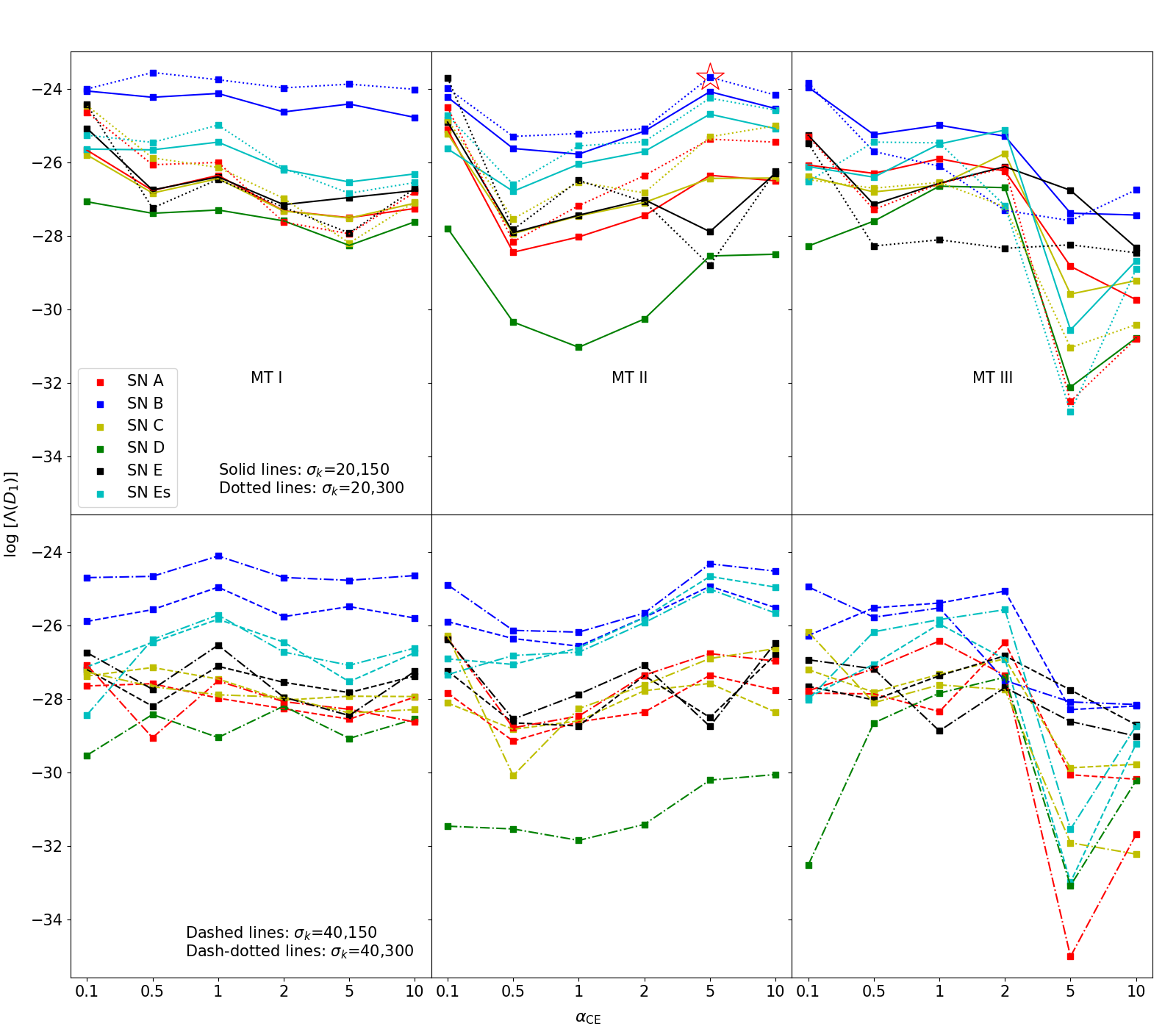}
\caption{Similar to Figure 5, but with the Doppler Shifting effect taken into account.}
\end{figure}


We then re-calculate $\Lambda (D_1)$  taking into account the Doppler shifting effect (see Table A), and present the results in Fig.~9. Comparing it with Fig.~5, we see the following features.
\begin{itemize}
     \item Considering a soft cut-off improves most of the models in general.
     \item Similar as in Fig.~5, MT I and II models are better than MT III model, and SN B model is more preferred compared with other SN models.
     \item In most cases, $\sigma_{\rm CCSN}=300$ km\,s$^{-1}$ and $\sigma_{\rm ECSN}=20$ km\,s$^{-1}$ are preferred over other kick velocity combinations.
\end{itemize}
In addition, we find that adopting the prescription proposed by \citet{Bray2018} for the kick velocity of USSN (dark lines) does not significantly change the properties of the DNS population. However, assuming that USSN kicks follow the same distribution as ECSN kicks (i.e., $\sigma_{\rm k}= 20/40$ km s$^{-1}$, cyan lines) can significantly improve the match between SN E models and observation.

Combining with Fig.~6, we find that among the ten best models selected by the Bayesian method, model MTIISNBk20300$\alpha$5 seems to best match the derived DNS merger rate.



\section{Discussion and Conclusions} \label{sec:highlight}

In this work, we construct 324 binary evolution models to investigate the formation of Galactic DNSs, considering the effects of MT stability, SN mechanisms and related kick velocity distribution, and CE efficiency. We evolve $10^8$ binary systems and obtain a few to several hundred newborn DNS systems in each model. Using the formulae of \citet{Peters1964}, we trace the orbital evolution of each DNS with time driven by GW radiation. Combining with empirical relations for known Galactic pulsars, we predict the radio characteristics and merger rates of the DNSs in the Galaxy. We aim to evaluate the influence of the key parameters by comparing theory with observation.

\subsection{CE evolution and MT stability}
The evolution of the CE plays a key role in the formation of DNSs, but  the specific physical processes are still not well understood \citep{Ivanova2013}.
Regarding the formation of DNS in the Galaxy, different groups have used different CE parameters in their studies.
\citet{Andrews2015} combined $\alpha_{\rm CE}$ with $\lambda$ and set $\alpha_{\rm CE}\lambda=0.01$, 0.05, 0.1, 0.2, 0.25, 0.3, and 1.0. Using the Bayesian analysis method, they concluded that $\alpha_{\rm CE}\lambda=0.3-1$ is consistent with pulsar observations, while $\alpha_{\rm CE}\lambda \leq 0.25$ can be effectively ruled out, with the best value being 0.5. Considering that the value of $\lambda$ for supergiants is less than 0.1 (Dewi \& Tauris 2000; Podsiadlowski et al. 2003; Wang et al. 2016), their results indicate that $\alpha_{\rm CE}$ should be $\gtrsim 5$. \citet{Vigna2018} used the values of $\lambda$  calculated by \citet{Xu2010a,Xu2010} and set $\alpha_{\rm CE}=0.1$, 1 and 10 to simulate the formation of Galactic DNSs. Their results show that the value of $\alpha_{\rm CE}$ mainly affects the birth rate and merger rate of DNSs, but the Bayesian analysis neither
clearly favours nor disfavours the $\alpha_{\rm CE}$ variations with respect
to the fiducial value $1$. \citet{Kruckow2018} calculated the binding energy parameter $\lambda$ of star's envelope based on their evolutionary state and set $\alpha_{\rm CE}=0.2$, 0.5, and 0.8, and showed that the birth rate and merger rate of DNSs increase with increasing $\alpha_{\rm CE}$. \citet{Chu2022} took $\lambda$ to be 0.5 and $\alpha_{\rm CE}=0.1$, 1, and 10. Their results of Bayesian analysis showed that the formation of Galactic DNSs requires a high efficiency factor ($\alpha_{\rm CE}=10$). \citet{Sgalletta2023} calculated the value of $\lambda$ based on the model given by \citet{Claeys2014}, and set $\alpha_{\rm CE}=0.5$, 1, 3, and 5.  Comparing their results with the predicted DNS merger rate \citep[$37^{+24}_{-11}$Myr$^{-1}$ in][]{Pol2020}, they concluded that only $\alpha_{\rm CE}=3$ matches the observation.


The occurrence of CE evolution is critically dependent on the stability of MT \citep{Soberman1997,Shao2014,Mandel2022}. \cite{Packet1981} pointed out that, when a star accretes about 10\%$\sim$15\% of its original mass, it can be spun up to the critical rotation, and the centrifugal force would inhibit further accretion \citep{Petrovic2005}. However, in close binaries tidal synchronization can prevent the star reaching the critical rotation \citep{deMin2013}. In addition, stars with strong magnetic fields  may decelerate their rotation under the effect of magnetic braking \citep{Dervi2010,Deschamps2013,Shao2014}.

Be/X-ray binaries are often used to evaluate the efficiency of stellar accretion \citep[e.g.,][]{deMin2013,Shao2014,Vinciguerra2020}. \citet{Shao2014} investigated the binary interaction channel for the formation of Be/X-ray binaries. They showed that, if adopting the rotation-limited accretion model (Model I), the masses of the accreting stars are mostly less than $6\,M_{\odot}$, which are inconsistent with the masses of the Be stars. \citet{Vinciguerra2020} compared the observed population of Be/X-ray binaries in the Small Magellanic Cloud  with simulated populations of Be/X-ray binaries. Their results suggest that at least 30 per cent  of the mass donated by the progenitor of the NS is typically accreted by the B-star companion. Additionally both \citet{Shao2014} and \citet{Vinciguerra2020} disfavor conservative MT. On the other hand, it has been suggested that the rotation-limited accretion model may help form the Galactic Wolf-Rayet/O binaries \citep{Shao2016}, Be/black hole binaries like MWC 656 with an orbital period of $\sim$ 60 days \citep{Casares2014}, and OB star-black hole system binaries \citep{Langer2020}, without the occurrence of CE evolution \citep{Podsiadlowski2003,Shao2014,Grudzinska2015}.


In this work, we adopt three MT models, the calculated $\lambda$ values by \citet{Xu2010,Xu2010a} and \citet{Wang2016}, and $\alpha_{\rm CE}=0.1-10$ to simulate the formation of DNS in the Galaxy. We find that the Pulsar-NS distribution does not effectively constrain the MT model and the value of $\alpha_{\rm CE}$, but the derived DNS merger rate prefers MT II model and $\alpha_{\rm CE}= 0.5\sim10$. However, our best fit models with and without considering the Doppler shifting effect reveal $\alpha_{\rm CE}=5$ and 0.5, respectively. The former is broadly compatible with the studies on the formation of black hole low-mass X-ray binaries \citep[e.g.,][]{Kalogera1999,Podsiadlowski2003,Kiel2006,Yungelson2008}, while the latter is consistent with studies on the formation of white dwarf binaries \citep[e.g.,][]{Zorotovic2010,Davis2012,Nandez2015,Ge2022,Zorotovic2022,Scherbak2022}. Since the progenitors of black holes and white dwarfs are respectively mass stars and intermediate-/low-mass stars, the difference could imply possible dependence of $\alpha_{\rm CE}$ on stellar mass and mass ratio.


\subsection{SN models and natal kicks}

In this work, we adopt rapid, delay and stochastic CCSN models to estimate the CO masses. We note that under the same initial conditions, there are no significant difference between the predicted PSR-NS populations in the rapid and delay SN models, while the performance of the stochastic SN model seems not be as well as the other two. The main cause is associated with its different prescriptions of the kick velocity distribution. There is also large difference between the DNS merger rates calculated with the rapid/delay models and with the stochastic SN models.

Besides CCSN, we also consider the contribution from ECSN and USSN.
We assume two possible helium core mass range $M_{\rm ecs}=(1.83-2.25)M_{\odot}$, and $(1.83-2.75)M_{\odot}$ for ECSN. Our simulations show that using $M_{\rm ecs}=(1.83-2.75)M_{\odot}$ better match the observations than using $M_{\rm ecs}=(1.83-2.25)M_{\odot}$, indicating that a large portion of NSs in DNSs have been formed via ECSN with a relatively small kick velocity \citep[see also][]{Shao2018b}.

We adopt Maxwellian distributions for the kick velocities of CCSN and ECSN.
For the specific combinations of the $\sigma_{\rm k}$ values (20/150, 40/150, 20/300, 40/300 km\,s$^{-1}$), our Bayesian analysis results indicate that $\sigma_{\rm k}=20/300$ km\,s$^{-1}$ is better than other combinations in most cases. Furthermore, if taking $\sigma_{\rm k}=300$ km\,s$^{-1}$ for CCSN, only MT II model can match the merger rate of Galactic DNSs.

As to USSN, it is not certain whether the NS is exerted a significantly smaller kick compared with CCSN. For example, \citet{Vigna2018} adopted a Maxwellian distribution with $\sigma_{\rm k}=30$ km\,s$^{-1}$ for USSN kick velocities following \citet{Podsiadlowski2004}, but numerical simulation by \cite{Schneider2021} suggested that stripped stars on average give rise to lower-mass NSs, higher explosion energies, and higher kick velocities ($\sigma_{\rm k}=315\pm 24$ km\,s$^{-1}$). In this work, we use the prescription proposed by \citet{Bray2018} for the kick velocity of USSN, and find that there are not significantly changes in the properties of the DNS population and their merger rate. If we instead assume that USSN kicks follow the same distribution as ECSN kicks (i.e., $\sigma_{\rm k}= 20/40$ km s$^{-1}$), then the match between SN E models and observation is greatly improved, as shown in Fig.~9 (in the cyan color and labeled as SN Es). Therefore, the USSN kick distribution will be an important subject for further research.

\subsection{Summary}

In this work we construct 324 models for the formation of Galactic DNSs taking into account various possible combinations of the key parameters. We try to explore their influence by comparing the properties of modeled DNS population with observation.
Our work differs from previous studies in several ways. For example, compared with \citet{Andrews2015} and \citet{Chu2022}, we used the  $\lambda$ values calculated from stellar evolution and considered the radio properties \footnote{We adopt beamming effect, death line, and doppler shifting to deal the pulsar select effect. Note that we do not account for CE accretion when fitting $P_{\rm s}$ and $B$ of the fort-born NS.} and merger rate of DNSs to constrain models. Our results show that the double-core CE channel (Channel II) also contribute to the formation of the Milky Way DNSs, which was not considered by \citet{Andrews2015}. Compared with \citet{Vigna2018} and \citet{Sgalletta2023}, we considered more choices for the MT efficiency. We systematically explore the effects of different MT models, CE efficiency factor $\alpha_{\rm CE}$, SNs and NS kicks on the formation of Galactic DNSs, and obtain useful estimates of the key parameters by comparing models predictions with both observed Galactic DNSs and their inferred merger rates. We list all the adopted models in Table 2 and use Bayesian analysis to evaluate them (see Section 3.2). Our main results are summarized as follows:

1. Comparing the predicted and derived Galactic DNSs merger rates reveals the allowed CE efficiency factor $\alpha_{\rm CE}$ ranging from 0.5 to 10. However, the Bayesian analysis indicates that there is no apparent dependence of the Pulsar-NS distribution on $\alpha_{\rm CE}$. We find that $\alpha_{\rm CE}=0.5$ and 5 in our best fit models when the Doppler Shifting effect is not considered and when a soft cut-off related to the Doppler Shifting effect is considered, respectively.

2. Adopting MT II model (with $50\%$ accretion efficiency) would allow more models to match the Galactic DNS merger rates, while Bayesian analysis indicates that there is no apparent dependence of the Pulsar-NS distribution on the MT mode.

3. Model SN B (with $M_{\rm ecs}=[1.83-2.75]\,M_{\odot}$ for ECSN) seems to significantly better match observation than other SN models, implying that ECSN plays an important role in the formation of Galactic DNSs.

4. In most cases, $\sigma_{\rm ECSN}=20$ km\,s$^{-1}$  and $\sigma_{\rm CCSN}=300$ km\,s$^{-1}$  combination are preferred compared with other kick velocity combinations. The results are also sensitively dependent on the kick prescriptions for USSN.

Considering the fact that the current DNS sample is rather limited and there are many parameters involved in the formation of DNSs, we caution that our results are preliminary and subject to several issues. For example, we assume that the Milky Way has a constant metal abundance and a constant SFR, which will affect the merger rate of DNSs, even though the impact on the current merger rate is limited according to previous researches \citep{Chu2022,Sgalletta2023}. We do not take into account the merger history of the Milky Way when calculating the 3D movement of DNSs, which might have significant influence on the spatial distribution of DNSs. In addition, we assume that part of the accreting material is ejected out of the binary in the form of isotropic wind, while the actual situation for mass loss may be more complicated, which impacts the orbital evolution \citep{Soberman1997,Kruckow2018,Vinciguerra2020}. Together with more detections of DNSs in the Milky Way,
a thorough investigation on the evolutionary history
of DNSs needs to incorporate the formation of massive binaries \citep[e.g.,][]{Wang2020}, high-mass X-ray binaries \citep[e.g.,][]{Misra2023}, and  binary pulsars in a self-consistent way. This will be the subject of the future study.

\begin{acknowledgements}
We thank an anonymous referee for valuable comments that helped improve the manuscript. This work was supported by the National Key Research and Development Program of China (2021YFA0718500), the Natural Science Foundation of Chian under grant No. 12041301, 12121003, 12203051.
\end{acknowledgements}

\bibliographystyle{aasjournal}
\bibliography{DNS}{}

\appendix

\section*{Appendix A: Results of Bayesian analysis}\label{secA1}

\startlongtable
\begin{deluxetable*}{c|cccc|cccc|cc}
\tablenum{A}
\tablecaption{The posterior probabilities, ranking numbers, birth rates and merger rates of Galactic DNSs in all adopted models. The second and third sets correspond to the results without and with considering the Doppler shifting effect, respectively.}
\tablehead{
Model &    log$\Lambda (\rm D_1)$  &  level$_1$ &    log$\Lambda (\rm D_2)$  &  level$_2$  & log$\Lambda (\rm D_1)$  &  level$_1$ &    log$\Lambda (\rm D_2)$  &  level$_2$  & Birth rates    &   Merger rates  \\
& \multicolumn{4}{c}{No Doppler shifting effect}  & \multicolumn{4}{c}{Doppler shifting effect}  &  ($\rm Myr^{-1}$) & ($\rm Myr^{-1}$) }
\startdata
MTIISNEk20150$\alpha$0.1 & -23.84 & 1 & -28.33 & 1 & -24.9 & 36 & -29.92 & 38 & 8.39 & 5.97 \\
\textbf{MTIISNBk20300$\alpha$0.5} & -23.9 & 2 & -28.4 & 2 & -25.29 & 52 & -29.94 & 39 & 55.79 & 44.06 \\
MTIISNAk20150$\alpha$0.1 & -24.25 & 3 & -29.61 & 7 & -25.1 & 45 & -30.09 & 47 & 9.33 & 6.58 \\
MTIISNBk40300$\alpha$0.5 & -24.54 & 4 & -30.21 & 16 & -26.13 & 84 & -31.91 & 99 & 32.72 & 27.35 \\
MTIISNAk20300$\alpha$0.1 & -24.69 & 5 & -29.24 & 5 & -24.49 & 23 & -28.77 & 18 & 5.55 & 3.92 \\
MTISNBk20150$\alpha$0.1 & -24.81 & 6 & -28.92 & 3 & -24.05 & 12 & -28.23 & 9 & 13.44 & 6.91 \\
MTIISNBk40300$\alpha$10 & -24.92 & 7 & -30.44 & 19 & -24.51 & 24 & -29.83 & 36 & 69.11 & 40.99 \\
MTIISNBk20150$\alpha$0.1 & -24.96 & 8 & -29.68 & 9 & -24.22 & 17 & -28.48 & 14 & 14.41 & 8.5 \\
MTIISNBk20150$\alpha$0.5 & -25.01 & 9 & -29.88 & 10 & -25.61 & 64 & -30.71 & 61 & 69.19 & 53.66 \\
MTIISNCk20150$\alpha$0.1 & -25.05 & 10 & -30.81 & 27 & -25.21 & 48 & -30.04 & 46 & 10.05 & 7.1 \\
MTIISNBk20300$\alpha$1 & -25.12 & 11 & -29.68 & 8 & -25.21 & 47 & -29.71 & 33 & 57.05 & 33.92 \\
MTIISNBk20300$\alpha$0.1 & -25.2 & 12 & -29.19 & 4 & -23.98 & 9 & -27.7 & 2 & 10.78 & 5.99 \\
MTIISNCk20300$\alpha$1 & -25.2 & 13 & -31.27 & 37 & -26.53 & 113 & -32.34 & 123 & 24.8 & 17.94 \\
MTIIISNBk20150$\alpha$0.1 & -25.24 & 14 & -29.54 & 6 & -23.95 & 7 & -28.11 & 8 & 11.92 & 6.47 \\
MTISNCk20300$\alpha$0.1 & -25.29 & 15 & -29.93 & 12 & -24.44 & 22 & -28.78 & 19 & 5.07 & 3.25 \\
MTIISNBk40300$\alpha$5 & -25.3 & 16 & -30.81 & 26 & -24.32 & 19 & -29.53 & 29 & 59.39 & 38.26 \\
MTIISNEk20300$\alpha$0.1 & -25.32 & 17 & -29.92 & 11 & -23.7 & 3 & -28.09 & 7 & 5.35 & 3.84 \\
MTIISNBk20150$\alpha$1 & -25.45 & 18 & -30.25 & 17 & -25.77 & 70 & -30.48 & 57 & 72.45 & 41.92 \\
\textbf{MTIISNBk20300$\alpha$5} & -25.47 & 19 & -29.96 & 13 & -23.67 & 2 & -27.77 & 4 & 90.84 & 47.63 \\
MTIISNAk20300$\alpha$5 & -25.48 & 20 & -29.98 & 14 & -25.36 & 55 & -30.44 & 54 & 26.24 & 17.89 \\
MTIIISNEk20300$\alpha$0.1 & -25.51 & 21 & -30.64 & 23 & -25.47 & 58 & -30.03 & 45 & 4.67 & 3.21 \\
MTISNBk40300$\alpha$0.5 & -25.54 & 22 & -31.3 & 39 & -24.66 & 29 & -30.1 & 48 & 10.94 & 6.29 \\
MTISNBk40300$\alpha$5 & -25.59 & 23 & -31.38 & 44 & -24.76 & 32 & -30.16 & 51 & 31.04 & 19.02 \\
MTIISNBk20150$\alpha$5 & -25.59 & 24 & -30.38 & 18 & -24.07 & 13 & -28.53 & 15 & 119.99 & 64.97 \\
MTIISNBk40150$\alpha$10 & -25.59 & 25 & -31.49 & 50 & -25.51 & 60 & -31.24 & 77 & 109.69 & 68.38 \\
MTIISNBk40150$\alpha$0.5 & -25.64 & 26 & -31.41 & 47 & -26.35 & 98 & -32.09 & 112 & 43.13 & 34.56 \\
MTISNCk20150$\alpha$0.1 & -25.74 & 27 & -30.45 & 20 & -25.79 & 73 & -30.83 & 65 & 9.85 & 6.22 \\
MTIIISNCk20150$\alpha$0.1 & -25.79 & 28 & -31.42 & 48 & -26.38 & 101 & -31.74 & 92 & 8.61 & 6.16 \\
MTISNEk20150$\alpha$0.1 & -25.83 & 29 & -31.03 & 31 & -25.07 & 43 & -29.85 & 37 & 8.51 & 5.57 \\
MTIISNBk40150$\alpha$5 & -25.83 & 30 & -31.69 & 54 & -24.93 & 37 & -30.45 & 56 & 81.95 & 52.75 \\
MTIIISNBk20300$\alpha$0.1 & -25.84 & 31 & -30.18 & 15 & -23.83 & 5 & -27.52 & 1 & 9.26 & 4.75 \\
MTISNBk40300$\alpha$10 & -25.88 & 32 & -31.35 & 42 & -24.63 & 28 & -29.97 & 42 & 35.51 & 19.77 \\
MTIISNBk20150$\alpha$10 & -25.89 & 33 & -30.92 & 28 & -24.53 & 25 & -29.14 & 22 & 140.2 & 68.71 \\
MTIISNCk40300$\alpha$0.1 & -25.92 & 34 & -32.5 & 77 & -26.27 & 92 & -32.03 & 109 & 3.99 & 3.15 \\
MTIISNCk20300$\alpha$5 & -25.94 & 35 & -31.35 & 43 & -25.29 & 54 & -30.44 & 55 & 28.88 & 20.1 \\
MTISNBk20150$\alpha$5 & -26.01 & 36 & -31.22 & 36 & -24.4 & 20 & -29.26 & 23 & 72.9 & 39.59 \\
MTISNBk40150$\alpha$1 & -26.02 & 37 & -31.55 & 51 & -24.95 & 39 & -30.43 & 53 & 22.13 & 11.46 \\
MTIISNCk20300$\alpha$10 & -26.03 & 38 & -31.1 & 34 & -25.0 & 41 & -29.77 & 34 & 27.96 & 17.99 \\
MTIIISNBk40150$\alpha$0.1 & -26.05 & 39 & -31.29 & 38 & -26.27 & 91 & -31.88 & 97 & 7.43 & 5.13 \\
MTIISNBk20300$\alpha$2 & -26.09 & 40 & -30.54 & 21 & -25.07 & 44 & -29.28 & 25 & 63.79 & 24.69 \\
MTIISNBk20150$\alpha$2 & -26.09 & 41 & -30.95 & 29 & -25.14 & 46 & -29.66 & 32 & 81.28 & 31.11 \\
MTIISNBk40150$\alpha$0.1 & -26.1 & 42 & -31.94 & 58 & -25.89 & 77 & -31.6 & 87 & 9.52 & 6.82 \\
MTIISNBk20300$\alpha$10 & -26.11 & 43 & -30.8 & 25 & -24.16 & 16 & -28.38 & 13 & 94.37 & 40.4 \\
MTISNBk40150$\alpha$0.5 & -26.11 & 44 & -32.06 & 65 & -25.56 & 63 & -31.15 & 74 & 20.54 & 11.48 \\
MTIISNBk40300$\alpha$2 & -26.13 & 45 & -31.8 & 56 & -25.65 & 65 & -30.91 & 66 & 36.59 & 16.37 \\
MTISNBk40150$\alpha$10 & -26.14 & 46 & -32.11 & 66 & -25.79 & 74 & -31.42 & 82 & 70.8 & 41.3 \\
MTISNBk20150$\alpha$10 & -26.19 & 47 & -31.31 & 40 & -24.77 & 33 & -29.55 & 31 & 86.86 & 42.52 \\
MTISNBk40300$\alpha$0.1 & -26.19 & 48 & -31.08 & 32 & -24.69 & 31 & -29.55 & 30 & 5.6 & 3.73 \\
MTISNBk20150$\alpha$0.5 & -26.22 & 49 & -31.56 & 52 & -24.22 & 18 & -28.62 & 16 & 27.43 & 12.56 \\
MTIISNAk40300$\alpha$5 & -26.23 & 50 & -32.11 & 67 & -26.76 & 130 & -32.82 & 147 & 20.77 & 16.15 \\
MTIISNCk20300$\alpha$0.1 & -26.23 & 51 & -30.72 & 24 & -24.83 & 34 & -29.49 & 28 & 5.57 & 3.88 \\
MTIISNEk20300$\alpha$2 & -26.24 & 52 & -31.32 & 41 & -27.06 & 145 & -32.3 & 120 & 28.52 & 18.32 \\
MTISNEk20300$\alpha$0.1 & -26.26 & 53 & -30.95 & 30 & -24.42 & 21 & -28.9 & 20 & 4.96 & 3.19 \\
MTIIISNBk40300$\alpha$0.1 & -26.27 & 54 & -31.62 & 53 & -24.94 & 38 & -29.96 & 41 & 4.98 & 3.46 \\
MTIISNEk20150$\alpha$0.5 & -26.28 & 55 & -32.31 & 71 & -27.9 & 225 & -34.0 & 208 & 39.42 & 32.75 \\
MTIISNEk20300$\alpha$10 & -26.32 & 56 & -31.08 & 33 & -26.24 & 90 & -30.97 & 67 & 35.1 & 21.44 \\
MTISNBk20300$\alpha$0.1 & -26.32 & 57 & -30.55 & 22 & -23.99 & 10 & -27.78 & 5 & 9.66 & 4.64 \\
MTIISNAk20300$\alpha$10 & -26.33 & 58 & -31.46 & 49 & -25.44 & 57 & -30.24 & 52 & 27.59 & 18.18 \\
MTISNBk40150$\alpha$5 & -26.34 & 59 & -32.41 & 72 & -25.48 & 59 & -31.18 & 75 & 58.86 & 37.2 \\
MTISNBk20150$\alpha$1 & -26.35 & 60 & -31.39 & 45 & -24.12 & 15 & -28.73 & 17 & 30.24 & 12.58 \\
MTIISNBk40150$\alpha$2 & -26.36 & 61 & -31.96 & 60 & -25.77 & 72 & -31.1 & 73 & 48.89 & 20.86 \\
MTIISNAk40300$\alpha$2 & -26.42 & 62 & -32.92 & 90 & -27.34 & 174 & -33.86 & 200 & 13.25 & 8.39 \\
MTIISNAk20150$\alpha$5 & -26.44 & 63 & -31.97 & 61 & -26.34 & 96 & -31.98 & 105 & 55.95 & 34.77 \\
MTIISNEk40300$\alpha$10 & -26.44 & 64 & -32.0 & 62 & -26.46 & 110 & -31.78 & 94 & 29.71 & 21.39 \\
MTIISNCk40300$\alpha$5 & -26.47 & 65 & -32.78 & 82 & -26.89 & 138 & -32.86 & 149 & 23.1 & 18.2 \\
MTIISNBk40300$\alpha$1 & -26.48 & 66 & -32.06 & 64 & -26.17 & 88 & -31.54 & 84 & 32.83 & 21.26 \\
MTIISNBk40150$\alpha$1 & -26.52 & 67 & -32.47 & 75 & -26.56 & 116 & -32.19 & 117 & 44.47 & 27.05 \\
MTISNAk20150$\alpha$0.1 & -26.6 & 68 & -32.18 & 70 & -25.66 & 66 & -30.72 & 62 & 9.39 & 5.78 \\
MTIISNAk20150$\alpha$10 & -26.6 & 69 & -32.02 & 63 & -26.49 & 112 & -32.17 & 116 & 75.21 & 49.15 \\
MTISNBk40300$\alpha$1 & -26.64 & 70 & -32.16 & 69 & -24.1 & 14 & -29.09 & 21 & 11.64 & 5.98 \\
MTISNAk40300$\alpha$0.1 & -26.69 & 71 & -32.41 & 73 & -27.07 & 148 & -32.74 & 141 & 3.96 & 2.99 \\
MTISNBk20300$\alpha$0.5 & -26.72 & 72 & -31.22 & 35 & -23.55 & 1 & -27.72 & 3 & 17.32 & 7.12 \\
MTIISNBk40300$\alpha$0.1 & -26.77 & 73 & -31.39 & 46 & -24.89 & 35 & -30.55 & 60 & 6.36 & 4.66 \\
MTISNBk40150$\alpha$2 & -26.79 & 74 & -32.92 & 91 & -25.75 & 69 & -31.42 & 81 & 32.19 & 14.48 \\
MTIISNCk40150$\alpha$5 & -26.82 & 75 & -33.53 & 109 & -27.57 & 195 & -33.99 & 207 & 54.32 & 39.87 \\
MTIISNCk40300$\alpha$2 & -26.82 & 76 & -32.88 & 87 & -27.61 & 202 & -33.39 & 174 & 14.06 & 9.1 \\
MTISNCk20150$\alpha$2 & -26.82 & 77 & -32.91 & 88 & -27.31 & 172 & -33.42 & 177 & 27.21 & 15.71 \\
MTIISNCk20150$\alpha$5 & -26.83 & 78 & -32.84 & 85 & -26.43 & 105 & -32.04 & 110 & 65.55 & 42.85 \\
MTIISNAk40300$\alpha$10 & -26.87 & 79 & -33.26 & 98 & -26.96 & 142 & -33.07 & 158 & 25.36 & 19.48 \\
MTIISNEk40150$\alpha$2 & -26.9 & 80 & -32.92 & 89 & -27.36 & 177 & -33.16 & 161 & 27.54 & 16.62 \\
MTISNEk20150$\alpha$1 & -26.92 & 81 & -32.5 & 76 & -26.39 & 102 & -32.11 & 113 & 17.78 & 11.25 \\
MTISNBk20300$\alpha$10 & -26.93 & 82 & -31.93 & 57 & -24.0 & 11 & -28.33 & 11 & 50.08 & 20.71 \\
MTIISNEk20300$\alpha$1 & -26.95 & 83 & -31.74 & 55 & -26.48 & 111 & -31.81 & 95 & 26.14 & 19.02 \\
MTIISNEk20150$\alpha$2 & -27.08 & 84 & -32.51 & 78 & -27.0 & 144 & -32.44 & 127 & 42.0 & 24.29 \\
MTIISNEk20300$\alpha$0.5 & -27.11 & 85 & -32.88 & 86 & -27.82 & 217 & -33.97 & 205 & 27.31 & 23.46 \\
MTISNEk40150$\alpha$1 & -27.14 & 86 & -33.68 & 116 & -27.1 & 152 & -32.86 & 152 & 15.78 & 10.77 \\
MTISNBk20300$\alpha$5 & -27.14 & 87 & -32.16 & 68 & -23.86 & 6 & -28.34 & 12 & 44.33 & 21.22 \\
MTISNBk20300$\alpha$1 & -27.15 & 88 & -31.94 & 59 & -23.74 & 4 & -27.86 & 6 & 19.39 & 7.02 \\
MTIISNAk20300$\alpha$1 & -27.15 & 89 & -33.6 & 113 & -27.17 & 156 & -32.79 & 144 & 23.51 & 16.41 \\
MTISNBk40300$\alpha$2 & -27.16 & 90 & -33.51 & 108 & -24.69 & 30 & -30.03 & 44 & 17.6 & 8.37 \\
MTIISNEk40300$\alpha$0.1 & -27.17 & 91 & -33.48 & 103 & -26.37 & 100 & -33.03 & 157 & 3.7 & 3.04 \\
MTIISNDk20$\alpha$0.1 & -27.18 & 92 & -33.59 & 111 & -27.78 & 214 & -34.09 & 215 & 12.98 & 9.22 \\
MTISNBk20150$\alpha$2 & -27.21 & 93 & -32.46 & 74 & -24.62 & 26 & -29.27 & 24 & 42.44 & 15.76 \\
MTIISNCk20300$\alpha$0.5 & -27.24 & 94 & -33.51 & 107 & -27.53 & 193 & -33.6 & 183 & 28.17 & 24.41 \\
MTISNCk40150$\alpha$10 & -27.28 & 95 & -33.84 & 120 & -27.93 & 228 & -34.39 & 232 & 51.26 & 34.27 \\
MTISNEk20150$\alpha$5 & -27.28 & 96 & -33.13 & 95 & -26.95 & 141 & -33.02 & 156 & 41.98 & 29.84 \\
MTISNCk20150$\alpha$10 & -27.29 & 97 & -33.68 & 117 & -27.1 & 151 & -33.16 & 160 & 53.86 & 34.62 \\
MTIISNCk40300$\alpha$10 & -27.29 & 98 & -33.27 & 99 & -26.62 & 120 & -32.58 & 134 & 25.42 & 19.16 \\
MTISNCk20150$\alpha$5 & -27.33 & 99 & -33.64 & 115 & -27.51 & 192 & -33.68 & 189 & 51.0 & 36.73 \\
MTIISNEk20150$\alpha$10 & -27.36 & 100 & -32.68 & 80 & -26.29 & 93 & -31.4 & 79 & 70.98 & 46.52 \\
MTISNEk20150$\alpha$10 & -27.38 & 101 & -33.49 & 104 & -26.76 & 131 & -32.55 & 131 & 50.14 & 35.57 \\
MTISNBk20300$\alpha$2 & -27.39 & 102 & -32.57 & 79 & -23.96 & 8 & -28.23 & 10 & 27.2 & 9.45 \\
MTIISNCk20150$\alpha$2 & -27.4 & 103 & -32.84 & 84 & -27.08 & 150 & -32.86 & 150 & 44.44 & 22.93 \\
MTISNEk40300$\alpha$1 & -27.41 & 104 & -33.01 & 92 & -26.53 & 114 & -32.39 & 125 & 6.51 & 4.96 \\
MTIISNEk20150$\alpha$1 & -27.41 & 105 & -33.16 & 96 & -27.43 & 185 & -33.3 & 168 & 40.56 & 27.23 \\
MTIISNAk40300$\alpha$0.1 & -27.41 & 106 & -34.83 & 162 & -26.34 & 95 & -32.33 & 122 & 3.86 & 3.1 \\
MTISNEk40150$\alpha$10 & -27.43 & 107 & -34.06 & 126 & -27.37 & 179 & -33.73 & 192 & 48.2 & 35.54 \\
MTIISNCk40150$\alpha$0.1 & -27.43 & 108 & -33.78 & 119 & -28.1 & 242 & -34.74 & 247 & 7.97 & 6.09 \\
MTIISNEk40150$\alpha$10 & -27.44 & 109 & -33.35 & 101 & -26.78 & 132 & -32.48 & 128 & 63.31 & 46.43 \\
MTIISNAk20300$\alpha$2 & -27.47 & 110 & -32.76 & 81 & -26.35 & 99 & -31.64 & 89 & 21.71 & 13.13 \\
MTISNAk20150$\alpha$5 & -27.48 & 111 & -33.49 & 106 & -27.5 & 190 & -33.8 & 195 & 47.08 & 33.46 \\
MTISNCk40300$\alpha$0.1 & -27.52 & 112 & -33.47 & 102 & -27.26 & 167 & -33.81 & 197 & 3.97 & 3.04 \\
MTIISNCk20150$\alpha$1 & -27.56 & 113 & -33.59 & 112 & -27.45 & 187 & -33.44 & 178 & 43.53 & 28.26 \\
MTIIISNEk20150$\alpha$10 & -27.58 & 114 & -34.13 & 128 & -28.31 & 258 & -35.26 & 276 & 34.19 & 28.6 \\
MTIISNEk40150$\alpha$0.5 & -27.58 & 115 & -34.26 & 135 & -28.64 & 279 & -35.11 & 272 & 26.21 & 22.1 \\
MTIISNCk20150$\alpha$10 & -27.62 & 116 & -33.27 & 100 & -26.42 & 104 & -31.96 & 103 & 77.88 & 48.92 \\
MTIISNCk20150$\alpha$0.5 & -27.62 & 117 & -34.31 & 137 & -27.93 & 229 & -34.35 & 229 & 44.76 & 37.0 \\
MTISNEk20300$\alpha$10 & -27.63 & 118 & -33.49 & 105 & -26.7 & 124 & -32.56 & 132 & 18.75 & 13.98 \\
MTIISNCk20300$\alpha$2 & -27.64 & 119 & -32.82 & 83 & -26.82 & 136 & -32.08 & 111 & 23.19 & 14.19 \\
MTISNEk40300$\alpha$10 & -27.64 & 120 & -33.71 & 118 & -27.24 & 165 & -33.18 & 162 & 16.86 & 13.68 \\
MTISNCk40150$\alpha$1 & -27.64 & 121 & -34.03 & 125 & -27.45 & 188 & -33.8 & 196 & 17.1 & 10.48 \\
MTISNAk20150$\alpha$1 & -27.68 & 122 & -33.91 & 123 & -26.34 & 97 & -32.19 & 118 & 18.03 & 10.7 \\
MTIIISNAk20300$\alpha$0.1 & -27.7 & 123 & -33.11 & 94 & -25.29 & 53 & -30.0 & 43 & 4.94 & 3.43 \\
MTISNAk40150$\alpha$10 & -27.71 & 124 & -34.22 & 133 & -27.94 & 231 & -34.68 & 244 & 52.4 & 36.55 \\
MTIISNEk40150$\alpha$0.1 & -27.72 & 125 & -34.7 & 158 & -27.23 & 163 & -32.89 & 153 & 6.35 & 4.95 \\
MTIISNAk20150$\alpha$0.5 & -27.75 & 126 & -34.15 & 129 & -28.44 & 265 & -34.88 & 255 & 42.89 & 35.11 \\
MTIISNEk40300$\alpha$1 & -27.76 & 127 & -33.96 & 124 & -27.86 & 222 & -33.88 & 202 & 15.9 & 11.97 \\
MTIIISNCk40150$\alpha$0.1 & -27.76 & 128 & -34.39 & 141 & -27.19 & 160 & -33.67 & 187 & 6.53 & 4.98 \\
MTIIISNAk40150$\alpha$0.5 & -27.77 & 129 & -34.54 & 150 & -27.85 & 221 & -34.09 & 213 & 6.05 & 3.11 \\
MTIISNEk20150$\alpha$5 & -27.79 & 130 & -33.19 & 97 & -27.88 & 223 & -33.54 & 182 & 58.53 & 37.62 \\
MTIIISNAk20150$\alpha$0.1 & -27.83 & 131 & -33.62 & 114 & -26.07 & 81 & -31.03 & 70 & 8.02 & 5.54 \\
MTIISNAk20150$\alpha$1 & -27.87 & 132 & -34.16 & 130 & -28.02 & 236 & -34.2 & 220 & 41.8 & 26.23 \\
MTISNEk40150$\alpha$5 & -27.87 & 133 & -34.63 & 154 & -27.82 & 216 & -34.24 & 222 & 39.9 & 29.52 \\
MTIIISNCk20150$\alpha$0.5 & -27.89 & 134 & -34.41 & 142 & -26.8 & 133 & -32.66 & 139 & 8.28 & 3.58 \\
MTIISNAk40300$\alpha$1 & -27.89 & 135 & -34.99 & 172 & -28.46 & 268 & -35.0 & 265 & 14.2 & 10.28 \\
MTIIISNEk40150$\alpha$10 & -27.92 & 136 & -34.98 & 170 & -28.69 & 281 & -35.38 & 280 & 32.13 & 27.88 \\
MTISNDk20$\alpha$5 & -27.92 & 137 & -34.44 & 143 & -28.25 & 250 & -34.65 & 243 & 36.9 & 27.44 \\
MTISNDk20$\alpha$10 & -27.93 & 138 & -34.32 & 138 & -27.61 & 201 & -33.65 & 186 & 37.4 & 22.56 \\
MTIISNAk40150$\alpha$10 & -27.94 & 139 & -34.88 & 166 & -27.75 & 210 & -34.32 & 226 & 69.12 & 50.59 \\
MTISNCk20150$\alpha$0.5 & -27.96 & 140 & -34.45 & 144 & -26.82 & 137 & -32.84 & 148 & 19.22 & 12.0 \\
MTISNCk40150$\alpha$5 & -28.01 & 141 & -34.66 & 155 & -27.92 & 227 & -34.6 & 240 & 48.52 & 36.14 \\
MTIISNCk40150$\alpha$0.5 & -28.01 & 142 & -35.12 & 181 & -28.82 & 286 & -35.37 & 279 & 30.11 & 24.79 \\
MTISNAk20150$\alpha$10 & -28.02 & 143 & -34.46 & 145 & -27.25 & 166 & -33.32 & 169 & 54.88 & 36.6 \\
MTIISNEk40300$\alpha$2 & -28.04 & 144 & -33.06 & 93 & -27.08 & 149 & -32.51 & 129 & 17.06 & 11.45 \\
MTIISNCk40150$\alpha$1 & -28.06 & 145 & -34.79 & 160 & -28.59 & 275 & -35.55 & 283 & 28.53 & 18.88 \\
MTIISNAk40150$\alpha$5 & -28.08 & 146 & -34.85 & 163 & -27.35 & 176 & -33.76 & 193 & 45.25 & 31.96 \\
MTIIISNEk20150$\alpha$0.1 & -28.1 & 147 & -33.89 & 121 & -25.26 & 50 & -30.55 & 59 & 6.86 & 4.69 \\
MTIIISNBk40150$\alpha$10 & -28.11 & 148 & -34.68 & 157 & -28.18 & 245 & -35.06 & 269 & 50.13 & 36.61 \\
MTIIISNDk40$\alpha$2 & -28.12 & 149 & -35.28 & 187 & -27.4 & 183 & -33.51 & 181 & 3.34 & 1.61 \\
MTISNEk40300$\alpha$0.1 & -28.13 & 150 & -34.76 & 159 & -26.73 & 125 & -32.02 & 108 & 3.72 & 2.92 \\
MTISNAk40150$\alpha$0.5 & -28.14 & 151 & -35.03 & 176 & -27.58 & 196 & -34.21 & 221 & 16.1 & 10.77 \\
MTISNDk40$\alpha$5 & -28.14 & 152 & -35.18 & 185 & -29.06 & 292 & -36.08 & 289 & 36.14 & 27.45 \\
MTISNEk40150$\alpha$0.5 & -28.17 & 153 & -34.46 & 146 & -28.19 & 248 & -34.58 & 238 & 15.53 & 11.17 \\
MTIISNCk40150$\alpha$10 & -28.17 & 154 & -34.97 & 168 & -28.35 & 261 & -34.97 & 262 & 71.2 & 50.35 \\
MTIIISNBk40150$\alpha$5 & -28.18 & 155 & -34.99 & 171 & -28.28 & 257 & -34.9 & 259 & 28.49 & 21.83 \\
MTIIISNBk40300$\alpha$10 & -28.18 & 156 & -35.53 & 199 & -28.14 & 243 & -35.15 & 274 & 29.45 & 22.06 \\
MTISNDk20$\alpha$0.1 & -28.18 & 157 & -34.52 & 149 & -27.06 & 146 & -32.39 & 124 & 9.03 & 4.81 \\
MTISNBk40150$\alpha$0.1 & -28.2 & 158 & -34.68 & 156 & -25.88 & 76 & -31.89 & 98 & 9.43 & 6.04 \\
MTISNAk40150$\alpha$0.1 & -28.22 & 159 & -33.54 & 110 & -27.64 & 204 & -34.36 & 230 & 8.04 & 5.48 \\
MTIISNAk40150$\alpha$2 & -28.23 & 160 & -35.17 & 184 & -28.35 & 262 & -34.62 & 241 & 27.22 & 14.38 \\
MTIISNAk40150$\alpha$0.5 & -28.27 & 161 & -35.29 & 188 & -29.14 & 293 & -36.12 & 290 & 28.09 & 23.13 \\
MTISNEk40150$\alpha$0.1 & -28.3 & 162 & -35.48 & 195 & -27.19 & 159 & -33.27 & 166 & 7.08 & 5.19 \\
MTISNAk20150$\alpha$0.5 & -28.3 & 163 & -34.33 & 139 & -26.76 & 129 & -32.65 & 136 & 17.88 & 11.27 \\
MTISNEk20150$\alpha$2 & -28.3 & 164 & -34.81 & 161 & -27.14 & 155 & -33.02 & 155 & 24.07 & 13.93 \\
MTISNEk20150$\alpha$0.5 & -28.31 & 165 & -34.59 & 153 & -26.74 & 127 & -32.33 & 121 & 17.49 & 11.67 \\
MTISNDk40$\alpha$2 & -28.32 & 166 & -35.79 & 213 & -28.19 & 246 & -34.56 & 237 & 15.2 & 7.26 \\
MTIISNCk40150$\alpha$2 & -28.32 & 167 & -35.07 & 177 & -27.77 & 212 & -34.11 & 216 & 29.3 & 15.85 \\
MTISNAk20300$\alpha$0.5 & -28.34 & 168 & -34.06 & 127 & -26.06 & 80 & -31.64 & 90 & 7.49 & 5.09 \\
MTISNCk20300$\alpha$0.5 & -28.34 & 169 & -34.16 & 131 & -25.87 & 75 & -31.1 & 72 & 8.03 & 5.42 \\
MTIIISNEk40150$\alpha$5 & -28.35 & 170 & -35.01 & 175 & -27.75 & 211 & -34.51 & 234 & 18.73 & 16.18 \\
MTIISNDk20$\alpha$5 & -28.36 & 171 & -35.15 & 182 & -28.54 & 271 & -34.82 & 251 & 55.22 & 40.63 \\
MTIISNEk20300$\alpha$5 & -28.4 & 172 & -33.9 & 122 & -28.79 & 285 & -34.27 & 223 & 40.68 & 25.66 \\
MTIISNAk20150$\alpha$2 & -28.41 & 173 & -34.59 & 152 & -27.44 & 186 & -33.29 & 167 & 41.61 & 20.99 \\
MTIIISNBk40150$\alpha$1 & -28.43 & 174 & -34.21 & 132 & -25.38 & 56 & -30.5 & 58 & 8.05 & 2.21 \\
MTIIISNBk40150$\alpha$0.5 & -28.43 & 175 & -34.37 & 140 & -25.51 & 61 & -30.78 & 63 & 8.64 & 2.9 \\
MTISNEk40150$\alpha$2 & -28.46 & 176 & -34.98 & 169 & -27.54 & 194 & -33.95 & 204 & 22.12 & 13.53 \\
MTISNCk20300$\alpha$5 & -28.5 & 177 & -34.55 & 151 & -28.19 & 247 & -34.59 & 239 & 18.47 & 14.17 \\
MTIISNAk20300$\alpha$0.5 & -28.53 & 178 & -34.87 & 165 & -28.15 & 244 & -34.18 & 219 & 26.72 & 23.03 \\
MTISNEk40300$\alpha$2 & -28.54 & 179 & -35.43 & 194 & -27.94 & 233 & -34.56 & 236 & 8.77 & 6.35 \\
MTIIISNAk40150$\alpha$0.1 & -28.54 & 180 & -35.0 & 173 & -27.83 & 220 & -34.28 & 224 & 6.09 & 4.63 \\
MTIISNAk40150$\alpha$0.1 & -28.57 & 181 & -36.41 & 233 & -27.83 & 218 & -35.04 & 266 & 7.37 & 5.6 \\
MTIISNEk40300$\alpha$5 & -28.61 & 182 & -35.01 & 174 & -28.74 & 283 & -34.84 & 253 & 29.9 & 21.15 \\
MTISNDk40$\alpha$10 & -28.63 & 183 & -35.69 & 204 & -28.54 & 272 & -35.33 & 277 & 36.63 & 22.8 \\
MTIISNEk40150$\alpha$1 & -28.64 & 184 & -35.15 & 183 & -28.72 & 282 & -35.06 & 270 & 26.44 & 18.32 \\
MTIIISNDk20$\alpha$10 & -28.66 & 185 & -35.61 & 201 & -30.77 & 309 & -37.86 & 306 & 30.0 & 22.26 \\
MTIIISNAk40300$\alpha$0.1 & -28.67 & 186 & -36.28 & 229 & -27.77 & 213 & -34.88 & 256 & 3.49 & 2.78 \\
MTISNAk20150$\alpha$2 & -28.7 & 187 & -35.11 & 180 & -27.31 & 171 & -33.32 & 170 & 24.5 & 13.0 \\
MTISNAk20300$\alpha$10 & -28.77 & 188 & -35.37 & 191 & -26.8 & 134 & -32.65 & 135 & 17.84 & 12.52 \\
MTISNAk40150$\alpha$5 & -28.8 & 189 & -36.11 & 224 & -28.54 & 273 & -35.46 & 282 & 44.8 & 33.09 \\
MTIIISNBk20150$\alpha$10 & -28.8 & 190 & -35.7 & 208 & -27.42 & 184 & -33.68 & 188 & 59.94 & 37.45 \\
MTISNCk40150$\alpha$0.5 & -28.8 & 191 & -35.97 & 219 & -27.14 & 154 & -33.41 & 176 & 17.65 & 11.62 \\
MTISNCk20300$\alpha$10 & -28.84 & 192 & -34.47 & 147 & -27.07 & 147 & -33.12 & 159 & 17.59 & 12.18 \\
MTISNDk40$\alpha$0.5 & -28.86 & 193 & -36.53 & 237 & -28.41 & 264 & -35.06 & 268 & 14.57 & 8.33 \\
MTISNDk40$\alpha$1 & -28.88 & 194 & -36.6 & 244 & -29.04 & 290 & -36.17 & 291 & 14.51 & 7.85 \\
MTISNEk20300$\alpha$2 & -28.89 & 195 & -34.86 & 164 & -27.21 & 162 & -33.2 & 164 & 10.35 & 6.74 \\
MTISNDk20$\alpha$2 & -28.91 & 196 & -35.09 & 178 & -27.58 & 198 & -33.36 & 172 & 16.07 & 7.16 \\
MTISNAk40300$\alpha$5 & -28.92 & 197 & -35.71 & 209 & -28.27 & 255 & -34.83 & 252 & 15.33 & 12.64 \\
MTIIISNCk20150$\alpha$1 & -28.93 & 198 & -35.61 & 202 & -26.61 & 119 & -32.16 & 115 & 7.03 & 2.64 \\
MTISNEk40300$\alpha$5 & -28.93 & 199 & -36.53 & 238 & -28.44 & 266 & -34.81 & 249 & 15.91 & 13.16 \\
MTISNAk40150$\alpha$1 & -28.97 & 200 & -36.56 & 240 & -27.97 & 234 & -34.48 & 233 & 16.1 & 10.23 \\
MTIIISNEk40150$\alpha$0.1 & -28.97 & 201 & -36.33 & 231 & -27.65 & 205 & -33.83 & 199 & 5.13 & 3.95 \\
MTIIISNCk20150$\alpha$2 & -28.98 & 202 & -35.09 & 179 & -25.75 & 68 & -31.04 & 71 & 6.38 & 2.4 \\
MTIISNCk40300$\alpha$1 & -28.98 & 203 & -35.64 & 203 & -28.25 & 251 & -35.13 & 273 & 15.01 & 11.24 \\
MTIISNDk20$\alpha$10 & -28.99 & 204 & -35.7 & 207 & -28.49 & 269 & -34.81 & 250 & 54.46 & 32.52 \\
MTISNCk40300$\alpha$5 & -29.0 & 205 & -35.72 & 210 & -28.37 & 263 & -35.84 & 287 & 16.5 & 13.76 \\
MTIISNDk20$\alpha$0.5 & -29.01 & 206 & -36.08 & 222 & -30.33 & 307 & -37.29 & 300 & 50.06 & 41.47 \\
MTISNDk20$\alpha$0.5 & -29.09 & 207 & -35.8 & 215 & -27.38 & 182 & -33.7 & 191 & 15.48 & 8.46 \\
MTIIISNAk40150$\alpha$2 & -29.13 & 208 & -35.96 & 218 & -26.46 & 108 & -32.66 & 138 & 4.62 & 2.18 \\
MTIIISNAk40150$\alpha$10 & -29.14 & 209 & -36.66 & 246 & -30.17 & 303 & -37.96 & 308 & 39.18 & 31.91 \\
MTIIISNDk20$\alpha$0.5 & -29.15 & 210 & -34.52 & 148 & -27.59 & 199 & -34.08 & 211 & 4.98 & 1.5 \\
MTIIISNAk40150$\alpha$5 & -29.15 & 211 & -36.58 & 241 & -30.05 & 301 & -38.13 & 310 & 23.42 & 20.4 \\
MTISNEk20300$\alpha$1 & -29.15 & 212 & -34.95 & 167 & -26.45 & 107 & -31.69 & 91 & 8.17 & 5.33 \\
MTISNEk20300$\alpha$5 & -29.16 & 213 & -35.49 & 196 & -27.91 & 226 & -33.97 & 206 & 17.9 & 13.67 \\
MTISNAk20300$\alpha$0.1 & -29.21 & 214 & -34.22 & 134 & -24.63 & 27 & -29.29 & 26 & 5.06 & 3.21 \\
MTISNAk20300$\alpha$5 & -29.22 & 215 & -36.09 & 223 & -27.93 & 230 & -34.32 & 227 & 17.5 & 13.07 \\
MTISNAk20300$\alpha$1 & -29.22 & 216 & -35.31 & 190 & -25.99 & 79 & -32.0 & 107 & 7.08 & 4.54 \\
MTISNCk40150$\alpha$2 & -29.23 & 217 & -36.67 & 247 & -28.02 & 237 & -34.94 & 261 & 25.0 & 15.3 \\
MTIISNAk40300$\alpha$0.5 & -29.24 & 218 & -36.58 & 243 & -28.78 & 284 & -36.22 & 293 & 16.56 & 14.6 \\
MTIIISNCk20300$\alpha$0.1 & -29.25 & 219 & -35.22 & 186 & -26.46 & 109 & -31.41 & 80 & 5.34 & 3.8 \\
MTIIISNEk20150$\alpha$5 & -29.28 & 220 & -35.39 & 192 & -26.75 & 128 & -32.72 & 140 & 22.26 & 18.39 \\
MTIIISNEk20150$\alpha$1 & -29.29 & 221 & -35.51 & 197 & -26.56 & 117 & -31.95 & 102 & 6.74 & 2.95 \\
MTIIISNDk20$\alpha$0.1 & -29.31 & 222 & -34.29 & 136 & -28.27 & 254 & -34.33 & 228 & 9.16 & 3.23 \\
MTIIISNEk40150$\alpha$0.5 & -29.31 & 223 & -35.58 & 200 & -28.02 & 235 & -34.89 & 258 & 6.14 & 3.5 \\
MTISNAk40150$\alpha$2 & -29.33 & 224 & -36.33 & 230 & -28.26 & 252 & -34.93 & 260 & 23.26 & 13.19 \\
MTIIISNBk40150$\alpha$2 & -29.34 & 225 & -35.7 & 206 & -25.06 & 42 & -30.12 & 49 & 8.04 & 2.16 \\
MTISNCk20150$\alpha$1 & -29.34 & 226 & -35.95 & 217 & -26.43 & 106 & -32.41 & 126 & 18.88 & 10.85 \\
MTISNCk40300$\alpha$10 & -29.37 & 227 & -36.84 & 251 & -28.28 & 256 & -34.98 & 264 & 15.73 & 12.21 \\
MTIISNEk40150$\alpha$5 & -29.38 & 228 & -36.47 & 234 & -28.49 & 270 & -34.86 & 254 & 45.3 & 32.18 \\
MTISNEk40300$\alpha$0.5 & -29.47 & 229 & -35.4 & 193 & -27.73 & 208 & -33.89 & 203 & 6.85 & 5.29 \\
MTISNAk40300$\alpha$10 & -29.54 & 230 & -36.95 & 254 & -28.62 & 278 & -35.4 & 281 & 15.72 & 12.38 \\
MTISNCk40300$\alpha$0.5 & -29.54 & 231 & -37.51 & 268 & -27.66 & 206 & -34.7 & 246 & 6.53 & 5.07 \\
MTIIISNBk40300$\alpha$0.5 & -29.57 & 232 & -35.89 & 216 & -25.77 & 71 & -31.0 & 68 & 5.29 & 1.81 \\
MTISNCk20300$\alpha$1 & -29.6 & 233 & -35.3 & 189 & -26.13 & 85 & -31.56 & 86 & 7.53 & 4.8 \\
MTISNCk40150$\alpha$0.1 & -29.62 & 234 & -37.08 & 257 & -27.37 & 180 & -33.86 & 201 & 8.4 & 5.84 \\
MTIIISNCk40150$\alpha$2 & -29.65 & 235 & -36.86 & 253 & -26.9 & 139 & -32.65 & 137 & 5.04 & 2.32 \\
MTIISNDk40$\alpha$0.5 & -29.71 & 236 & -37.85 & 281 & -31.53 & 315 & -38.85 & 313 & 46.89 & 39.16 \\
MTIIISNEk40150$\alpha$2 & -29.72 & 237 & -35.8 & 214 & -26.82 & 135 & -32.19 & 119 & 4.59 & 2.19 \\
MTIIISNEk40300$\alpha$10 & -29.72 & 238 & -38.86 & 293 & -29.0 & 289 & -36.61 & 296 & 15.94 & 14.41 \\
MTISNCk20300$\alpha$2 & -29.77 & 239 & -35.76 & 212 & -26.97 & 143 & -32.56 & 133 & 10.46 & 6.89 \\
MTISNAk40300$\alpha$2 & -29.8 & 240 & -36.58 & 242 & -28.06 & 238 & -34.89 & 257 & 8.57 & 6.14 \\
MTIIISNCk40150$\alpha$10 & -29.8 & 241 & -37.44 & 266 & -29.77 & 298 & -36.82 & 299 & 40.06 & 31.6 \\
MTIIISNAk20150$\alpha$5 & -29.82 & 242 & -36.84 & 252 & -28.82 & 287 & -35.62 & 285 & 28.09 & 23.54 \\
MTIIISNBk40300$\alpha$5 & -29.84 & 243 & -37.05 & 256 & -28.07 & 239 & -34.97 & 263 & 18.87 & 14.5 \\
MTIIISNEk40150$\alpha$1 & -29.84 & 244 & -36.2 & 226 & -27.36 & 178 & -33.19 & 163 & 5.28 & 2.76 \\
MTIISNCk40300$\alpha$0.5 & -29.94 & 245 & -37.31 & 263 & -30.08 & 302 & -37.57 & 303 & 17.43 & 15.47 \\
MTIIISNCk40150$\alpha$1 & -29.96 & 246 & -37.73 & 279 & -27.32 & 173 & -33.64 & 185 & 5.38 & 2.46 \\
MTIISNAk40150$\alpha$1 & -29.96 & 247 & -37.03 & 255 & -28.61 & 277 & -35.36 & 278 & 27.2 & 17.56 \\
MTISNAk40300$\alpha$1 & -29.97 & 248 & -37.96 & 283 & -27.5 & 191 & -34.75 & 248 & 5.65 & 4.27 \\
MTIIISNEk20150$\alpha$0.5 & -29.97 & 249 & -36.37 & 232 & -27.14 & 153 & -32.78 & 143 & 7.61 & 3.69 \\
MTISNAk20300$\alpha$2 & -30.01 & 250 & -36.48 & 235 & -27.62 & 203 & -34.05 & 210 & 10.19 & 6.3 \\
MTIISNEk40300$\alpha$0.5 & -30.01 & 251 & -37.59 & 270 & -28.54 & 274 & -35.06 & 267 & 16.75 & 14.89 \\
MTIIISNCk20150$\alpha$5 & -30.03 & 252 & -37.24 & 262 & -29.57 & 296 & -36.29 & 294 & 31.95 & 27.04 \\
MTISNCk40300$\alpha$2 & -30.06 & 253 & -37.08 & 259 & -27.94 & 232 & -34.31 & 225 & 8.78 & 6.56 \\
MTIISNDk40$\alpha$5 & -30.06 & 254 & -37.96 & 282 & -30.2 & 304 & -37.78 & 305 & 52.56 & 40.35 \\
MTIIISNAk20150$\alpha$10 & -30.07 & 255 & -37.68 & 276 & -29.73 & 297 & -36.62 & 297 & 42.27 & 32.26 \\
MTISNDk20$\alpha$1 & -30.07 & 256 & -36.74 & 249 & -27.29 & 169 & -33.37 & 173 & 15.53 & 8.02 \\
MTIIISNAk20150$\alpha$0.5 & -30.09 & 257 & -36.22 & 227 & -26.29 & 94 & -31.98 & 104 & 7.52 & 3.36 \\
MTIIISNCk20150$\alpha$10 & -30.09 & 258 & -37.18 & 260 & -29.21 & 294 & -36.21 & 292 & 43.53 & 32.3 \\
MTISNDk40$\alpha$0.1 & -30.1 & 259 & -37.32 & 265 & -29.52 & 295 & -36.75 & 298 & 8.05 & 4.79 \\
MTIISNDk40$\alpha$10 & -30.13 & 260 & -37.65 & 273 & -30.05 & 300 & -37.65 & 304 & 52.55 & 32.96 \\
MTIIISNBk40300$\alpha$1 & -30.14 & 261 & -36.56 & 239 & -25.52 & 62 & -30.8 & 64 & 4.87 & 1.15 \\
MTIIISNDk20$\alpha$1 & -30.14 & 262 & -36.16 & 225 & -26.64 & 121 & -31.93 & 101 & 3.57 & 0.96 \\
MTIISNDk20$\alpha$1 & -30.14 & 263 & -37.08 & 258 & -31.02 & 311 & -38.06 & 309 & 51.21 & 29.24 \\
MTIIISNAk20300$\alpha$1 & -30.16 & 264 & -36.04 & 220 & -26.57 & 118 & -32.16 & 114 & 2.97 & 1.13 \\
MTIIISNBk20150$\alpha$0.5 & -30.17 & 265 & -35.69 & 205 & -25.24 & 49 & -29.78 & 35 & 13.84 & 3.06 \\
MTISNEk20300$\alpha$0.5 & -30.19 & 266 & -36.08 & 221 & -27.23 & 164 & -32.81 & 146 & 8.57 & 5.78 \\
MTIIISNAk40300$\alpha$1 & -30.21 & 267 & -37.19 & 261 & -26.42 & 103 & -31.62 & 88 & 1.89 & 1.05 \\
MTIIISNBk20150$\alpha$1 & -30.21 & 268 & -35.73 & 211 & -24.98 & 40 & -29.48 & 27 & 13.83 & 2.46 \\
MTIIISNCk40150$\alpha$5 & -30.25 & 269 & -38.15 & 286 & -29.86 & 299 & -37.34 & 301 & 26.66 & 23.25 \\
MTIIISNBk20150$\alpha$5 & -30.38 & 270 & -36.65 & 245 & -27.37 & 181 & -33.5 & 180 & 38.84 & 25.39 \\
MTIIISNAk20150$\alpha$1 & -30.4 & 271 & -35.51 & 198 & -25.89 & 78 & -31.48 & 83 & 6.46 & 2.45 \\
MTIIISNBk20300$\alpha$10 & -30.42 & 272 & -37.49 & 267 & -26.74 & 126 & -32.8 & 145 & 38.38 & 23.41 \\
MTIIISNEk20300$\alpha$10 & -30.44 & 273 & -37.63 & 271 & -28.45 & 267 & -34.52 & 235 & 17.69 & 14.93 \\
MTISNCk40300$\alpha$1 & -30.44 & 274 & -38.14 & 285 & -27.88 & 224 & -34.64 & 242 & 6.09 & 4.61 \\
MTIIISNAk40300$\alpha$0.5 & -30.65 & 275 & -37.72 & 278 & -27.17 & 157 & -34.11 & 217 & 2.59 & 1.64 \\
MTIIISNAk40150$\alpha$1 & -30.71 & 276 & -37.71 & 277 & -28.34 & 260 & -34.09 & 214 & 5.01 & 2.29 \\
MTIIISNEk40300$\alpha$5 & -30.74 & 277 & -38.67 & 291 & -28.61 & 276 & -35.17 & 275 & 11.46 & 10.49 \\
MTIIISNDk20$\alpha$2 & -30.75 & 278 & -36.71 & 248 & -26.68 & 122 & -31.83 & 96 & 4.01 & 1.6 \\
MTIIISNBk20300$\alpha$1 & -30.75 & 279 & -36.49 & 236 & -26.09 & 82 & -31.0 & 69 & 10.15 & 1.21 \\
MTIIISNCk40300$\alpha$2 & -30.76 & 280 & -39.59 & 300 & -27.74 & 209 & -35.91 & 288 & 1.74 & 0.93 \\
MTIIISNCk20300$\alpha$5 & -30.76 & 281 & -36.28 & 228 & -31.03 & 312 & -38.58 & 311 & 17.82 & 15.55 \\
MTIIISNEk20300$\alpha$0.5 & -31.09 & 282 & -37.32 & 264 & -28.27 & 253 & -33.83 & 198 & 4.26 & 2.18 \\
MTIIISNAk20150$\alpha$2 & -31.11 & 283 & -37.54 & 269 & -26.23 & 89 & -31.98 & 106 & 5.9 & 2.13 \\
MTIIISNEk20150$\alpha$2 & -31.13 & 284 & -37.67 & 274 & -26.11 & 83 & -31.24 & 76 & 5.96 & 2.29 \\
MTISNAk40300$\alpha$0.5 & -31.19 & 285 & -37.64 & 272 & -29.04 & 291 & -35.08 & 271 & 6.14 & 4.81 \\
MTIIISNCk20300$\alpha$0.5 & -31.21 & 286 & -38.51 & 290 & -26.69 & 123 & -32.78 & 142 & 4.14 & 2.01 \\
MTIISNDk20$\alpha$2 & -31.26 & 287 & -38.31 & 287 & -30.25 & 306 & -36.61 & 295 & 43.39 & 14.12 \\
MTIIISNEk40300$\alpha$0.5 & -31.27 & 288 & -37.79 & 280 & -27.17 & 158 & -33.63 & 184 & 3.08 & 2.01 \\
MTIIISNCk20300$\alpha$10 & -31.32 & 289 & -38.98 & 295 & -30.4 & 308 & -37.5 & 302 & 20.15 & 15.72 \\
MTIIISNCk40300$\alpha$1 & -31.45 & 290 & -36.78 & 250 & -27.61 & 200 & -34.15 & 218 & 2.08 & 1.12 \\
MTIIISNAk20300$\alpha$5 & -31.46 & 291 & -38.94 & 294 & -32.49 & 321 & -40.22 & 319 & 16.54 & 14.31 \\
MTIIISNBk20150$\alpha$2 & -31.48 & 292 & -37.67 & 275 & -25.28 & 51 & -29.95 & 40 & 14.38 & 2.23 \\
MTIIISNAk40300$\alpha$10 & -31.54 & 293 & -39.61 & 301 & -31.68 & 316 & -40.32 & 320 & 17.05 & 14.66 \\
MTIIISNCk40150$\alpha$0.5 & -31.68 & 294 & -38.46 & 289 & -27.81 & 215 & -34.01 & 209 & 6.65 & 3.39 \\
MTIIISNDk40$\alpha$10 & -31.73 & 295 & -40.03 & 305 & -30.21 & 305 & -37.89 & 307 & 29.09 & 22.41 \\
MTIIISNDk20$\alpha$5 & -31.75 & 296 & -39.72 & 303 & -32.1 & 319 & -40.33 & 321 & 26.32 & 22.84 \\
MTIIISNAk20300$\alpha$10 & -31.76 & 297 & -39.82 & 304 & -30.78 & 310 & -38.64 & 312 & 19.24 & 15.1 \\
MTIIISNAk20300$\alpha$0.5 & -31.84 & 298 & -38.08 & 284 & -27.28 & 168 & -32.86 & 151 & 3.72 & 1.82 \\
MTIIISNEk40300$\alpha$0.1 & -31.92 & 299 & -39.06 & 297 & -26.93 & 140 & -33.0 & 154 & 3.16 & 2.59 \\
MTIISNDk40$\alpha$1 & -31.93 & 300 & -40.11 & 306 & -31.84 & 317 & -40.04 & 317 & 49.24 & 28.82 \\
MTIIISNDk40$\alpha$1 & -32.01 & 301 & -40.35 & 309 & -27.83 & 219 & -34.38 & 231 & 2.96 & 0.94 \\
MTIIISNCk40300$\alpha$0.5 & -32.1 & 302 & -39.11 & 299 & -28.09 & 240 & -34.09 & 212 & 2.84 & 1.81 \\
MTIIISNAk40300$\alpha$2 & -32.19 & 303 & -39.07 & 298 & -27.35 & 175 & -33.48 & 179 & 1.59 & 0.82 \\
MTIIISNBk20300$\alpha$2 & -32.4 & 304 & -38.45 & 288 & -27.29 & 170 & -31.93 & 100 & 10.9 & 1.06 \\
MTIIISNDk40$\alpha$0.5 & -32.42 & 305 & -40.34 & 308 & -28.64 & 280 & -35.76 & 286 & 4.4 & 1.54 \\
MTIISNDk40$\alpha$2 & -32.89 & 306 & -40.98 & 311 & -31.4 & 313 & -39.17 & 314 & 39.44 & 12.99 \\
MTIIISNEk40300$\alpha$2 & -32.91 & 307 & -38.98 & 296 & -27.68 & 207 & -33.21 & 165 & 1.95 & 1.03 \\
MTIIISNBk20300$\alpha$5 & -32.93 & 308 & -39.71 & 302 & -27.58 & 197 & -33.69 & 190 & 28.08 & 17.24 \\
MTIIISNCk40300$\alpha$10 & -32.95 & 309 & -41.29 & 314 & -32.21 & 320 & -40.04 & 318 & 17.5 & 14.95 \\
MTIIISNBk20300$\alpha$0.5 & -32.97 & 310 & -38.81 & 292 & -25.69 & 67 & -30.14 & 50 & 10.12 & 1.87 \\
MTIIISNDk40$\alpha$0.1 & -33.15 & 311 & -42.87 & 319 & -32.51 & 322 & -40.01 & 316 & 8.38 & 3.19 \\
MTIIISNEk20300$\alpha$1 & -33.33 & 312 & -40.75 & 310 & -28.1 & 241 & -33.4 & 175 & 3.55 & 1.53 \\
MTIIISNCk40300$\alpha$0.1 & -33.35 & 313 & -40.15 & 307 & -26.16 & 87 & -31.78 & 93 & 3.69 & 2.96 \\
MTIISNDk40$\alpha$0.1 & -33.55 & 314 & -41.24 & 312 & -31.45 & 314 & -39.37 & 315 & 12.31 & 9.3 \\
MTIIISNDk40$\alpha$5 & -33.57 & 315 & -44.22 & 322 & -33.07 & 323 & -42.51 & 323 & 25.08 & 22.35 \\
MTIIISNAk20300$\alpha$2 & -33.85 & 316 & -41.27 & 313 & -26.14 & 86 & -31.35 & 78 & 2.77 & 0.86 \\
MTIIISNBk40300$\alpha$2 & -34.02 & 317 & -41.5 & 316 & -27.49 & 189 & -33.34 & 171 & 5.01 & 0.98 \\
MTIIISNEk20300$\alpha$2 & -34.12 & 318 & -41.39 & 315 & -28.33 & 259 & -33.79 & 194 & 3.02 & 1.08 \\
MTIIISNEk40300$\alpha$1 & -34.66 & 319 & -42.27 & 318 & -28.86 & 288 & -35.55 & 284 & 2.35 & 1.37 \\
MTIIISNEk20300$\alpha$5 & -34.7 & 320 & -41.92 & 317 & -28.23 & 249 & -34.69 & 245 & 14.42 & 12.39 \\
MTIIISNCk40300$\alpha$5 & -35.35 & 321 & -44.97 & 323 & -31.9 & 318 & -40.34 & 322 & 13.82 & 12.76 \\
MTIIISNCk20300$\alpha$2 & -35.46 & 322 & -44.1 & 321 & -27.2 & 161 & -32.52 & 130 & 2.97 & 1.02 \\
MTIIISNCk20300$\alpha$1 & -35.86 & 323 & -43.16 & 320 & -26.54 & 115 & -31.55 & 85 & 3.17 & 1.24 \\
MTIIISNAk40300$\alpha$5 & -36.48 & 324 & -45.88 & 324 & -35.0 & 324 & -45.94 & 324 & 12.51 & 11.56 \\
\enddata
\tablecomments{The first set of data $D_1$ consists 15 fully confirmed DNSs, and the second set $D_2$ includes all 18 DNSs. Level$_1$ and level$_2$ of each model when ordered by log$\Lambda(\rm D_1)$ and log$\Lambda(\rm D_2)$. The best fit model in level$_1$ is displayed in boldface.
}

\end{deluxetable*}

\end{document}